\documentclass[12pt,preprint]{aastex}
\usepackage{lineno}

\usepackage{color}
\usepackage{lscape}
\usepackage{multirow}

\newcommand{\xmm}{{\em XMM-Newton}}

\newcommand{\sw}{{\em Swift}}

\newcommand{\msp}{3FGL\, J2039.6$-$5618}

\shorttitle{Multi-wavelength observations of a Fermi binary millisecond pulsar candidate}
\shortauthors{Salvetti et al.}

\begin{document}

\title{Multi-wavelength observations of 
\msp: a candidate redback millisecond pulsar}

\author{
D. Salvetti\altaffilmark{1},
R. P. Mignani\altaffilmark{1,2},
A. De Luca\altaffilmark{1,3},
C. Delvaux\altaffilmark{4},
C. Pallanca\altaffilmark{5},
A. Belfiore\altaffilmark{1},
M. Marelli\altaffilmark{1},
A. A. Breeveld\altaffilmark{6},
J. Greiner\altaffilmark{4},
W. Becker\altaffilmark{4},
D. Pizzocaro\altaffilmark{1,7}
}

\affil{\altaffilmark{1} INAF - Istituto di Astrofisica Spaziale e Fisica Cosmica Milano, via E. Bassini 15, 20133, Milano, Italy}
\affil{\altaffilmark{2} Janusz Gil Institute of Astronomy, University of Zielona G\'ora, Lubuska 2, 65-265, Zielona G\'ora, Poland}
\affil{\altaffilmark{3} Istituto Nazionale di Fisica Nucleare, Sezione di Pavia, Via Bassi 6, I-27100 Pavia, Italy}
\affil{\altaffilmark{4} Max-Planck Institut f\"ur Extraterrestrische Physik, Giessenbachstrasse 1, 85741 Garching bei M\"unchen, Germany}
\affil{\altaffilmark{5} Dipartimento di Fisica e Astronomia, Universit\`a degli Studi di Bologna, Viale Berti Pichat 6-2, I-40127, Bologna, Italy}
\affil{\altaffilmark{6} Mullard Space Science Laboratory, University College London, Holmbury St. Mary, Dorking, Surrey, RH5 6NT, UK}
\affil{\altaffilmark{7} Universit\'a degli Studi dell'Insubria, Via Ravasi 2, 21100 Varese, Italy}

\begin{abstract}
We present multi-wavelength observations of the unassociated $\gamma$-ray source \msp\ detected by the {\em Fermi} Large Area Telescope. The source $\gamma$-ray properties suggest that it is a pulsar, most likely a millisecond pulsar, for which neither radio nor $\gamma$-ray pulsations have been detected yet. We observed \msp\ with  \xmm\ and discovered several candidate X-ray counterparts within/close to the $\gamma$-ray error box. The brightest of these X-ray sources is variable with a period of 0.2245$\pm$0.0081 d. Its X-ray spectrum can be described by a power law with photon index $\Gamma_X =1.36\pm0.09$,
and hydrogen column density $N_{\rm H} < 4 \times 10^{20}$ cm$^{-2}$,  which gives an unabsorbed 0.3--10 keV X-ray flux of $1.02 \times 10^{-13}$ erg cm$^{-2}$ s$^{-1}$. Observations with the Gamma-Ray Burst Optical/Near-Infrared Detector (GROND) discovered an optical counterpart to this X-ray source, with a time-averaged magnitude $g'\sim 19.5$.  The counterpart features a flux modulation with a period of 0.22748$\pm$0.00043 d that coincides, within the errors, with that of the X-ray source, confirming the association based on the positional coincidence.  We interpret the observed X-ray/optical periodicity as the orbital period of a close binary system where one of the two members is a neutron star. The light curve profile of the companion star, with two asymmetric peaks, suggests that the optical emission comes from two regions at different temperatures on its tidally-distorted surface.
Based upon its X-ray and optical properties, we consider this source as  the most likely X-ray counterpart to \msp, which we propose to be a new redback system.  \end{abstract}


\section{Introduction}

The launch of the  {\em Fermi} Gamma-ray Space Telescope in June 2008  marked  a new era in $\gamma$-ray astronomy, thanks to the unprecedented performance of its Large Area Telescope (LAT;  Atwood et al.\ 2009). The recently released Third {\em Fermi}-LAT  $\gamma$-ray source catalogue (3FGL; Acero et al.\ 2015)  derived from the first 4 years of observations contains 3033 sources.
About 70\% of these sources have been either directly identified, e.g. either from the detection of $\gamma$-ray pulsations (pulsars) or correlated $\gamma$-ray and optical/radio variability (Active Galactic Nuclei, Novae, X-ray binaries), 
or associated with objects that are either known or potential $\gamma$-ray emitters.
The remaining 30\% of the 3FGL sources have not been associated with any object yet, hence they are referred to as {\em unassociated}, and their nature is unknown. 

Being pulsars the largest family of  $\gamma$-ray sources identified in the Galaxy ($\sim 160$ and counting\footnote{https://confluence.slac.stanford.edu/display/GLAMCOG/Public+List+of+LAT-Detected+Gamma-Ray+Pulsars}), a significant fraction of the  unassociated {\em Fermi}-LAT sources might be $\gamma$-ray pulsars. Some of them might have no or extremely faint radio emission, and escaped detection in all radio pulsar surveys so far.  Indeed, many of such radio-quiet (RQ), or radio faint (RF), $\gamma$-ray pulsars  have been discovered through blind periodicity searches in the $\gamma$-ray data (e.g., Abdo et al.\ 2009) thanks to the use of novel search techniques (e.g., Atwood et al.\ 2006; Pletsch et al.\ 2013).   About 45\% of the  $\gamma$-ray pulsars discovered by the {\em Fermi}-LAT are milli-second pulsars (MSPs). Interestingly, the vast majority of these MSPs ($\sim$ 80\%) are in binary systems. Some of them have an He white dwarf (WD)  companion star of mass  $0.1 M_{\odot} \la M_{\rm C} \la 0.5 M_{\odot}$, whereas others  have a usually non-degenerate companion  (a late main sequence star or a brown dwarf) which is ablated by irradiation from the pulsar wind.  Two distinct families of binary MSPs are recognized depending on the degree of the ablation processes: the Black Widow (BW) MSPs, where the companion is a very low-mass star of $M_{\rm C} \la 0.1 M_{\odot}$ almost fully ablated by the pulsar wind, and  the redback (RB) MSPs, where the companion is only partially ablated and has an higher mass of $M_{\rm C}\sim0.1$--$0.4 M_{\odot}$ (Roberts 2013). 

The use of automatic classification codes (e.g., Ackermann et al.\ 2012; Lee et al.\ 2012; Mirabal et al.\ 2012) based on the $\gamma$-ray characteristics is crucial to single out pulsar candidates among the many unassociated {\em Fermi}-LAT sources and optimise a systematic search for new $\gamma$-ray pulsars. Since $\gamma$-ray pulsars are also identified in the optical and X rays (e.g., Abdo et al.\ 2013), multi-wavelength follow-ups of unidentified {\em Fermi}-LAT sources are still key to confirm the proposed pulsar classifications, though. In particular, optical  observations are an important aid in the search for binary MSPs, for which blind periodicity searches in $\gamma$ rays  must account for the unknown orbital parameters, requiring a massive use of super-computing power facilities (Pletsch \& Clark 2014).  Indeed, optical observations yielded the identification of  the two {\em Fermi}-LAT sources 2FGL\, J2339.7$-$0531 and 2FGL\, J1311.7$-$3429 as binary MSPs prior to the detection of radio or $\gamma$-ray pulsations (Ray et al. in preparation; Pletsch et al.\ 2012; Ray et al.\ 2013) through the discovery of orbital modulations in the flux of their companion stars (Romani \& Shaw 2011; Kong et al.\ 2012; Romani et al.\ 2012; Kataoka et al.\ 2012).  In a similar way, new binary MSP candidates have been identified for the two unassociated {\em Fermi}-LAT sources 2FGL\, J1653.6$-$0159 (Romani et al.\ 2014; Kong et al.\ 2014) and 2FGL\, J0523.3$-$2530 (Strader et al.\ 2014).

As a part of a pilot project aimed at identifying different classes of  unassociated {\em Fermi}-LAT sources,
we studied  \msp.  This is a moderately bright $\gamma$-ray source (detection significance $\sim~25\sigma$) that was listed in both the First (Abdo et al.\ 2010) and Second (Nolan et al.\ 2012) {\em Fermi}-LAT  $\gamma$-ray source catalogues (a.k.a. 1FGL\, J2039.4$-$5621 and 2FGL\, J2039.8$-$5620, respectively), but has remained unassociated ever since.  
%
The \msp\ field was observed in X rays for the first time with the \sw/X-Ray Telescope (XRT) during snapshot observations (1 and 3.6 ks exposure times) but no candidate X-ray counterpart was detected within the  2FGL $\gamma$-ray source error circle (Takeuchi et al.\ 2013). In radio, no potential counterpart was found in the Sydney University Molonglo Sky Survey (SUMMS) source catalogue (Mauch et al.\ 2003) and in dedicated observations of unassociated 2FGL sources with the Australia Telescope Compact Array (Petrov et al.\  2013) and the Parkes radio telescope (Camilo et al.\ 2015).  At very high energies, \msp\ is not associated with any known TeV source\footnote{http://tevcat.uchicago.edu/}.
Based upon its $\gamma$-ray characteristics, \msp\  was classified as a very likely pulsar candidate by Mirabal et al.\ (2013).
Our newly developed classification method (Salvetti et al.\ 2013; Salvetti et al., in preparation) confirms that  \msp\  is a very likely pulsar candidate and suggests that it is, most probably, an MSP, either isolated or in a binary system (see Section~\ref{ann}).  However, since $\sim 80\%$ of the MSPs detected by the {\em Fermi}-LAT are in binary systems, one can expect that
would be a binary MSP.

We investigated this scenario through a multi-wavelength observation campaign 
(X rays, ultraviolet, optical, infrared) of \msp\ carried out  using both dedicated observations and exploiting data available in public archives.  The observations and data reduction are described in Section 2 and the results are presented in Section 3. Discussion and conclusion follow in Section 4 and 5, respectively.

\section{Observations and Data Reduction}

\subsection{Target selection}\label{ann}

Recently, we developed an advanced classification code (Salvetti et al.\ 2013) that can recognise different classes of $\gamma$-ray pulsars, e.g. young/middle-aged pulsars and MSPs. This code uses a statistical predictive method based on Artificial Neural Network (ANN) techniques to quantify the probability of a given source to be  MSP-like
on the basis of its $\gamma$-ray temporal and spectral characteristics.
The method is based on an advanced hierarchical ANN architecture consisting of 
 simple neural networks applied in sequence
  to discriminate pulsar-like from AGN-like objects in first place 
  and disentangle MSPs from young/middle age pulsars in second place. Such a method
  correctly classifies 84\% of the identified MSPs, 
while the false positive fraction is lower than 10\% (Salvetti et al., in preparation). We then applied the optimized hierarchical neural network to all {\em unassociated} 3FGL sources, to rank them according to their MSP probability function. As a result, \msp\ was classified as a MSP-like object with a probability greater than 99\%. Therefore, it stands out as an obvious candidate for multi-wavelength investigations. 

\subsection{X-ray observations}

We carried out an \xmm\ observation of the \msp\ $\gamma$-ray error box (Programme ID: 0720750301), which started on 2013 October 10 at 09:43:18 UT (revolution 2534) and lasted 44.6 ks. The pn detector (Struder et al.\ 2001) of the European Photon Imaging Camera (EPIC) instrument was operated in Extended Full Frame mode, with a time resolution of 200 ms over a $26\arcmin \times 27\arcmin$ Field-of-View (FoV), while the Metal Oxide Semi-conductor (MOS) detectors (Turner et al.\ 2001) were set in Full Frame mode (2.6 s time resolution on a 15$\arcmin$ radius FoV). The thin optical filter was used for the pn while a medium one was used for the MOS cameras. We retrieved the Observation Data Files (ODF) from the \xmm\ Science Archive\footnote{http://xmm.esac.esa.int/xsa/} and used the most recent release of the \xmm\ Science Analysis Software (SAS) v14.0 to analyze them. We performed a standard data processing, using the {\tt epproc} and {\tt emproc} tools, and screening for high particle background time intervals (e.g., De Luca et al.\ 2005). Our analysis revealed no significant contamination from soft protons. After the standard data processing, the good, dead-time corrected exposure time was  41.6 ks for the pn and 43.2 ks for the two MOS detectors.

\subsection{Ultraviolet, optical, infrared observations}\label{opt_uv}

In the optical/near-infrared (near-IR), we observed the \msp\ field with the Gamma-Ray Burst Optical/Near-Infrared Detector (GROND; Greiner et al.\ 2008) at the MPI/ESO 2.2m telescope on La Silla (Chile). 
The field was repeatedly observed 
on August 16, 17, and 18, 2014 
in the
g$'$, r$'$, i$'$, z$'$ bands in the optical
and in the J, H, K$_s$ bands  in the near-IR.   The observations were split into sequences of 18, 18, and 17 exposures per day, each consisting of four 115 s dithered exposures in the optical and forty eight 10 s dithered exposures in the near-IR.  
The observations were executed in  grey time with airmass between 1.12 and 1.28 and mean seeing of 1.0\arcsec. Single dithered exposures were reduced (bias subtraction, flat-fielding,  distortion correction) and stacked 
using standard {\sc IRAF}\footnote{IRAF is distributed by the National Optical Astronomy Observatories, which are operated by the Association of Universities for Research in Astronomy, Inc., under cooperative agreement with the National Science Foundation.} tasks  implemented in the GROND pipeline (Kr\"uhler et al.\ 2008; Yoldas et al.\ 2008).  The astrometry calibration was computed on single exposures against stars selected from the USNO-B1.0  catalogue (Monet et al.\ 2003) in the optical bands and the 2MASS catalogue (Skrutskie et al.\ 2006) in the  near-IR bands, 
yielding an accuracy of 0\farcs3 with respect to the chosen reference frame. The photometric calibration in the optical was computed  against a close-by  field from the Sloan  Digital Sky Survey (York et al.\ 2000) at $\delta=-10^{\circ}$ observed in the first night under photometric conditions. From the calibrated images we extracted a grid of
secondary photometric calibrators for direct on--the--frame calibration on the subsequent nights.  In the near-IR, the photometric calibration was 
computed 
against  2MASS stars identified  in the GROND field of view. The accuracy of the absolute photometry calibration was 0.02 magnitudes in  the g$'$, r$'$, i$'$, z$'$ bands, 0.03 magnitudes in J and H, and 0.05 in the K$_s$ band.

In addition to GROND, we used serendipitous JHK$_{\rm s}$ images of the \msp\ field taken on July 16 and 25 2010 with the near-IR camera (VIRCAM; Dalton et al.\ 2006) of the Visible and Infrared Survey Telescope for Astronomy (VISTA; Emerson et al.\ 2006).
The data set, processed and calibrated at the Cambridge Astronomical Survey Unit (CASU\footnote{http://casu.ast.cam.ac.uk/}),
consists of three sequences of 16 consecutive exposures of  15 s each in the J band and of 7.5 s each in both the H and K$_{\rm s}$ bands. 
In the optical and near-ultraviolet (near-UV) we used U, UVW1 ($\lambda=2910$ \AA;  $\Delta \lambda=1180$\AA) and UVM2 ($\lambda=2310$ \AA;  $\Delta \lambda=710$\AA) images from the \xmm\ Optical Monitor (OM; Mason et al.\ 2001) obtained in parallel to our observations, with exposure times of 1900, 2700, and 3080 s, respectively.
We also used archival UVW2 ($\lambda=2055$ \AA;  $\Delta \lambda=557$\AA) images from the \sw\ UltraViolet and Optical Telescope (UVOT; Roming et al.\ 2005)
performed on February 21 2011 (OBSID=00041479002), consisting of five exposures  
for  a total integration time of 3587 s.
The OM and UVOT data were processed and calibrated using the SAS tool {\tt omichain} and the {\sc HEASOFT} software package, respectively.

\section{Data analysis and results}

To search for possible counterparts of \msp\ in our multi-wavelength observations, we used as a reference its recent 3FGL position (Acero et al.\ 2015): $\alpha =20^{\rm h}  39^{\rm m} 40\fs32$ and  $\delta  = -56^\circ 18\arcmin 43\farcs6$ (J2000). Its associated 95\% confidence position error ellipse has semi-major and semi-minor axis of 2\arcmin.6 and 2\arcmin.4, respectively, and a position angle of $74\fdg95$, measured East of North. 
We started from our \xmm\ observations to find potential X-ray counterparts to {\msp}. Then, we used our multi-wavelength data base to single out those that are most likely associated with this $\gamma$-ray source. In particular, since we expect that \msp\ is 
a binary MSP, we focused our analysis on X-ray sources that show variability and/or have variable optical counterparts, possibly featuring periodic flux modulations.

\subsection{X-ray data analysis}

\subsubsection{Source detection}

\begin{figure}
\plotone{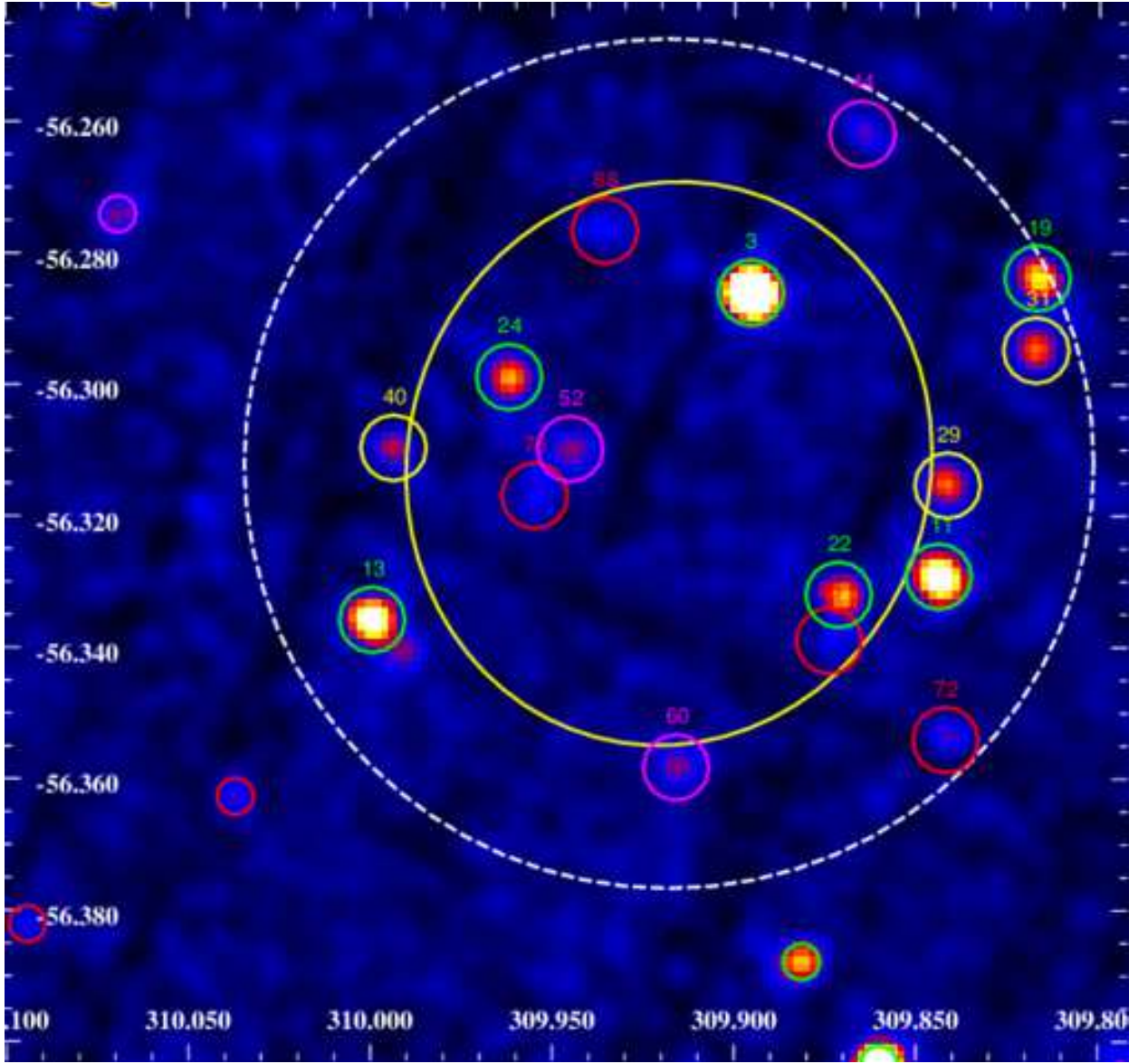}
\caption{0.3--10 keV exposure-corrected \xmm\ image of the \msp\ field obtained combining the images of the EPIC-pn camera and the two MOS detectors. The image has been smoothed using a Gaussian filter with a kernel radius of 3\arcsec. The 95\% confidence error ellipse of \msp\ is plotted in yellow. The white dashed ellipse corresponds to the 95\% error ellipse with the axis increased by 50\%.  X-ray sources detected within this region are highlighted with a circle of 18\arcsec\ radius 
and labeled as in Table 1, whereas other X-ray sources detected in the FoV are plotted with a radius of 10\arcsec. The colour of the circles correspond to the likelihood of the source detection (DET\_ML): $DET\_ML<25$ (red),  $25<DET\_ML<50$ (magenta),  $50<DET\_ML<100$ (yellow), $DET\_ML>100$ (green). \label{XMM-image}}
\end{figure}

For our X-ray analysis, we selected only 0$-$4 pattern events from the pn and 0$-$12 from the two MOS detectors  with the default flag {\em mask}. The source detection  in the 0.3--10 keV energy range was run simultaneously on the event lists of each of the EPIC-pn and MOS detectors using a maximum likelihood fitting with the SAS task {\tt edetect\_chain}
invoking other SAS tools to produce background, sensitivity, and vignetting-corrected exposure maps. 
The final source list includes 90 X-ray sources from both the pn and MOS detectors, with a combined pn+MOS detection likelihood greater than 10, corresponding to a significance above $ 3.5\sigma$.
Figure~\ref{XMM-image} shows the 0.3--10 keV exposure-corrected \xmm\ FoV obtained combining the images of the EPIC-pn and MOS detectors. We focused our analysis on the 16  X-ray sources detected within, or close to, the 95\% confidence position error ellipse of \msp. 
We summarized the positions, spectral parameters, fluxes, and variability indices of these sources in Table~\ref{tab1}.

\subsubsection{Spectral analysis}\label{spectra}

For each EPIC detector we extracted the source 
photons using an extraction radius of 20\arcsec, while we extracted background photons from source-free regions in the same CCD chip as the source, with radii of 50\arcsec--120\arcsec.
For each detector, we used the SAS task {\tt specgroup} to rebin all the extracted spectra and have at least 25 counts for each background-subtracted spectral channel
and generated ad hoc response matrices and ancillary files using the SAS tasks {\tt rmfgen} and {\tt arfgen}.
For each source, we fitted simultaneously the pn and MOS spectra using {\sc XSPEC} v12.8, forcing the same set of parameters and considering three different spectral models: a {\em power-law} (PL), well suited for both AGN and pulsars, an {\em apec} (AP) for stellar coronae, and a {\em black-body} (BB) for the pulsar thermal component. In all cases, the hydrogen column density $N_{\rm H}$ was left as a free parameter. For each emission model we computed the 90\% confidence level error on the spectral parameters. When the best-fit $N_{\rm H}$ values were comparable to zero, 
we assumed the measured uncertainties to determine the $3\sigma$ confidence level upper limit. Spectra with very low counts were fitted after fixing the $N_{\rm H}$ to the estimated value along the line of sight ($5\times10^{20}$ cm$^{-2}$; Dickey \& Lockman 1990), or fixing either the photon index or the temperature to  typical values, i.e. $\Gamma_{\rm X} =2$ and  $kT=3.5$ keV or $kT=0.2$ keV in case of an AP and a BB model, respectively.

\begin{figure}
\plotone{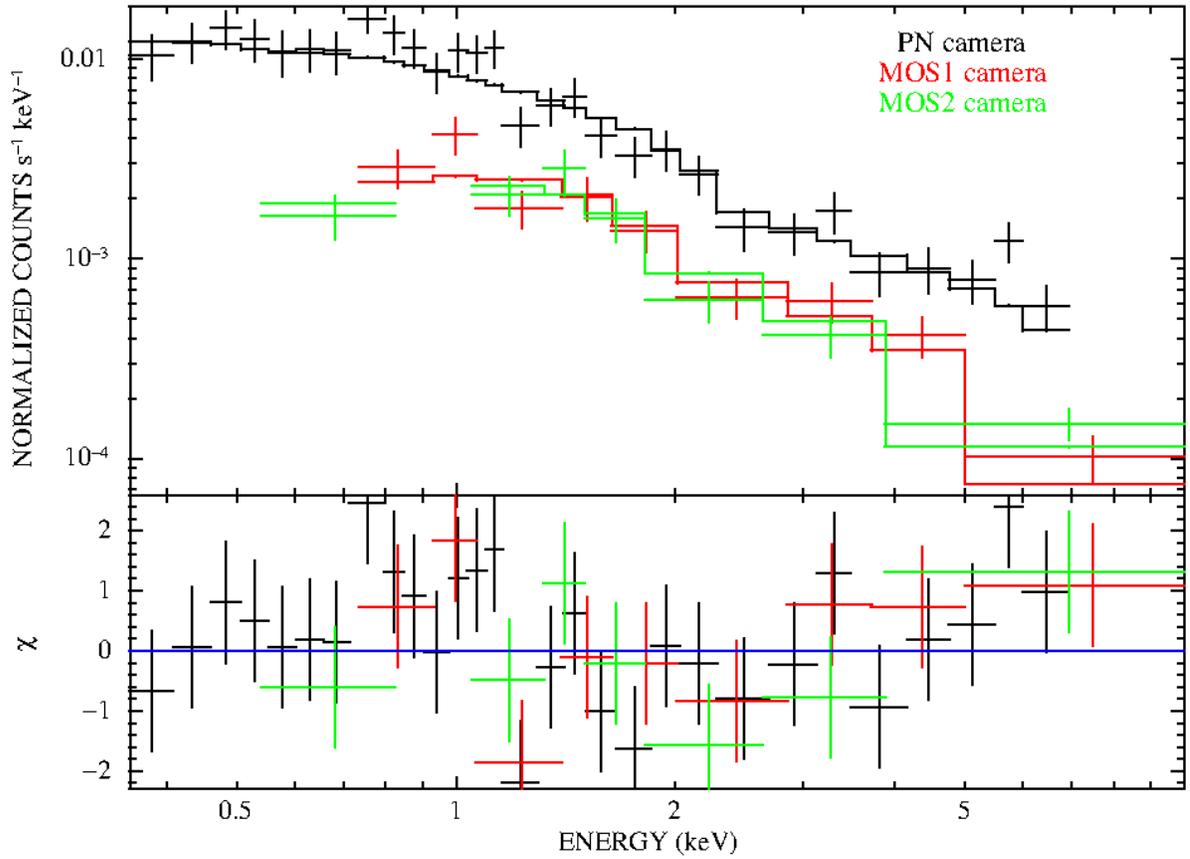}
\caption{Binned spectrum of Source 3, the brightest source among the most probable candidate X-ray counterparts to  \msp,  obtained with each of the EPIC detectors and best-fit with PL models. 
\label{spec_src3}}
\end{figure}

As shown in Table~\ref{tab1}, only Source 3 and 11, the two brightest X-ray sources in our sample, were best fitted by a simple PL model, with a $\chi^2$=52.32 (42 degrees of freedom, d.o.f.) for the former and $\chi^2$=16.06 (22 d.o.f.) for the latter. For the remaining 13 sources, at least two different models were required to obtain an acceptable fit with null hypothesis probability $>$ 0.01. Only for Source 88 it was not possible to obtain acceptable results with any of the selected spectral models. Indeed, this source is the faintest in our sample and 
is likely spurious. Figure~\ref{spec_src3} shows the binned spectrum of Source 3
extracted simultaneously from each of the EPIC detectors, together with the best-fit PL model. Since most MSPs are characterized by a significant thermal component to their X-ray spectra, we tried to fit the  spectra of Source 3 and Source 11 with an absorbed BB plus PL model. This improved the accuracy of the fit only for Source 3, with a $\chi^2$=34.9 (40 d.o.f.). 
From the F-test (Bevington 1969),  we computed a 0.0003 probability that adding a thermal component to the model would produce a chance improvement to the fit.  Since this probability corresponds to a $\sim$3.6$\sigma$ significance only,
hereafter we ignored the absorbed BB plus PL spectral model. 
Since for most X-ray sources the measured counts are too few to clearly discriminate among different spectral models, we checked whether we could extract qualitative spectral information from an hardness ratio (HR) analysis (Marelli et al.\  2014). However, we found that the observed HRs are compatible with different spectral models and, therefore, are not constraining.

\subsubsection{Variability analysis}\label{x_var}

\begin{figure}
\plottwo{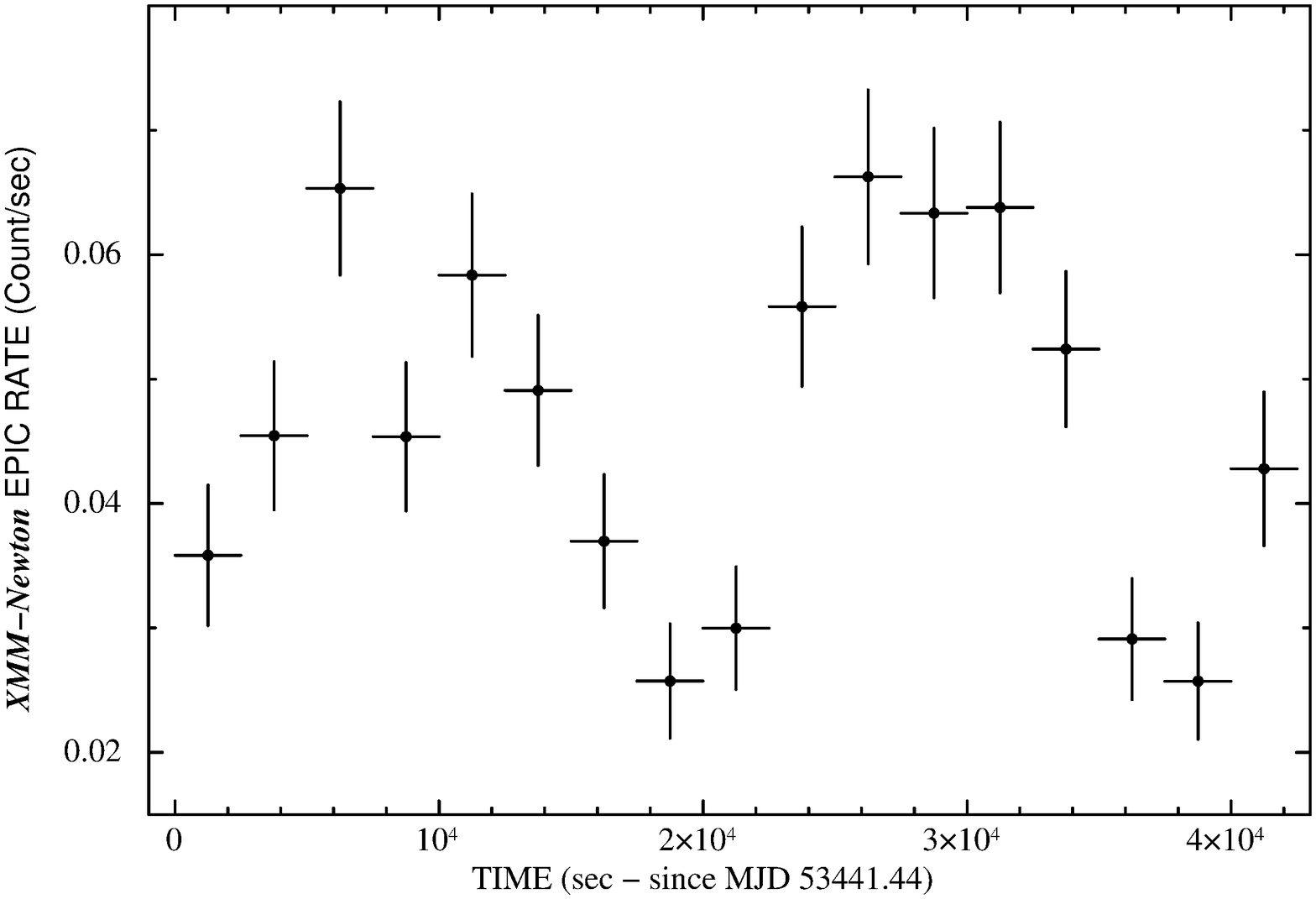}{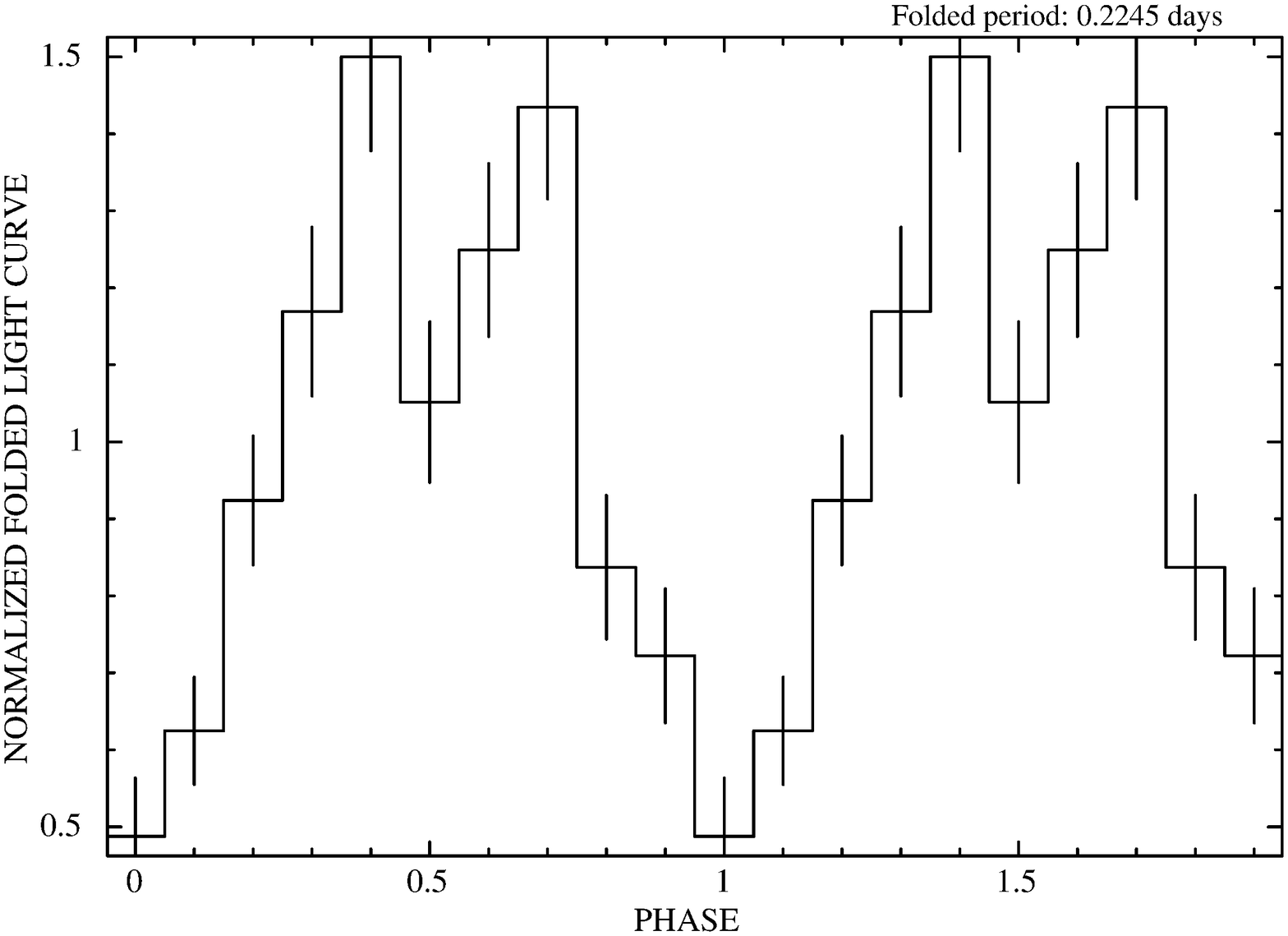}
\caption{{\em Left}: Background-subtracted light curve combining data from the 3 EPIC cameras for Source 3 in the 0.3--10 keV energy range, sampled with a bin time of 2500 s. 
{\em Right}:  Same but folded around the best period of 0.2245 days normalized to the average source intensity. In both panels, error bars are reported at 1$\sigma$.\label{lc_src3}}
\end{figure}

In order to detect possible time variability during the \xmm\ observation, we generated standard light curves from the pn-data
for all the 16 X-ray sources 
in Table 1. Starting from the source and background regions described in Section~\ref{spectra}, we extracted  source+background and 
background light curves, respectively,
with the SAS task {\tt evselect} and combined them into a background-subtracted light curve with the task {\tt epiclccorr}. This task also corrects the time series for 
vignetting, bad pixels, chip gaps, quantum efficiency, dead time, exposure and good time intervals. For each source we generated a light curve with time binning of 2500 s, or multiple of this value, to have at least 25 counts per bin and we ran a $\chi^2$ test to evaluate the variability significance.  Only Source 3 is characterized by a significant variability during the observation ($\chi^2=66.18$, with 16 d.o.f.), with a chance probability of $4.61\times10^{-8}$ ($>5\sigma$). After combining the data from all the three EPIC cameras the Source 3 light curve shows a more apparent variability (Figure~\ref{lc_src3}, left), with a chance probability of $7.43\times10^{-14}$, whereas the other X-ray still show no evidence of significant variability.

The Source 3 light curve 
also hints at a possibly periodic modulation.
We converted photon arrival times to the Solar system Barycentric Dynamical Time (TBD) with the SAS task {\tt barycen} and used the FTOOL task {\tt efsearch} to find the best period in the light curve through a maximum $\chi^2$ test. 
We folded the light curve with periods ranging from 100 to 43000 s, the latter comparable to the length of the \xmm\ observation.   We found the best-fit period  at 0.2245$\pm$0.0081 d, where the period uncertainty was obtained following Leahy (1987). The corresponding $\chi^2$ of 104.5 (9 d.o.f.) gives a chance probability of $6.38\times10^{-14}$,  accounting for the number of trials, and makes the periodicity statistically significant ($\sim 7.5~\sigma$).  The presence of a periodic signal at the corresponding frequency of $\sim 5\times10^{-5}$ Hz was independently confirmed by the power spectrum produced with the FTOOL {\tt powspec}. We note that the best-fit X-ray period of Source 3 is comparable to about half the length of the \xmm\ observation, so that 
the observed periodicity might be spurious. We examined the light curves of other comparably bright X-ray sources detected in the whole \xmm\ FoV and found that
none of them showed evidence of periodicity at any time scales. Nonetheless, the fact that the length of the \xmm\ observation only covers  $\sim$ 2.3 cycles, prevents us to firmly claim that Source 3 is periodic. 
The  X-ray light curve folded at the best-fit period (Figure~\ref{lc_src3}, right) is characterised by two peaks separated in phase by 0.31$\pm$0.04.
Therefore, it cannot be described by a simple sinusoidal or Gaussian model, with a null hypothesis probability lower than $10^{-3}$.  
We checked whether the folded X-ray light curve of Source 3 varied as a function of the energy and whether the X-ray spectrum changed as a function of the phase. In both cases, however, the available statistics is not sufficient to highlight significant differences in the energy-resolved light curves and the phase-resolved spectra.

We looked for archival X-ray images of the \msp\ field to  check for long-term variability of Source 3. The field was  serendipitously observed with the  X-ray Imaging Spectrometer (XIS) of {\em Suzaku} (Mitsuda et al.\ 2007)  on October 28, 2010 for a total exposure time of 21.5 ks 
(OBSID 705028010).  
We extracted source counts from a circle of radius 1\farcm3 around the best \xmm\ coordinates of Source 3 and
background counts from a nearby, source-free 2\arcmin-radius circle
using the HEASOFT(v.6.16) tool {\tt xselect}
and summed the spectra 
from all the XIS cameras with the {\tt mathpha}, {\tt addarf} and {\tt addrmf} tools. 
We obtained 350 counts, of which $\sim$50\% are from the source.
A fit with a PL gives a null hypothesis probability of 0.06 (4 d.o.f.), with $N_{\rm H}<5\times10^{21}$ cm$^{-2}$ (90\% upper limit), photon index $\Gamma$=1.3$\pm$0.4 and an unabsorbed flux in the 0.3--10 keV energy range of 1.7$_{-1.0}^{+0.3}$ $\times10^{-13}$ erg cm$^{-2}$ s$^{-1}$, fully compatible with the \xmm\ results.
The \xmm\ count-rate and spectral parameters (Table~\ref{tab1}) are also compatible with the non-detection above the 3$\sigma$ threshold ($\sim$ 1.05 $\times10^{-13}$ erg cm$^{-2}$ s$^{-1}$) of Source 3 in the short \sw/XRT images of Takeuchi et al.\  (2013), taken in 2010 and 2011. Therefore, we find no evidence of variability of Source 3 on time scales of three years.

\subsection{IR/Optical/UV analysis}

\subsubsection{Source cross-identification}

\begin{figure}
\plotone{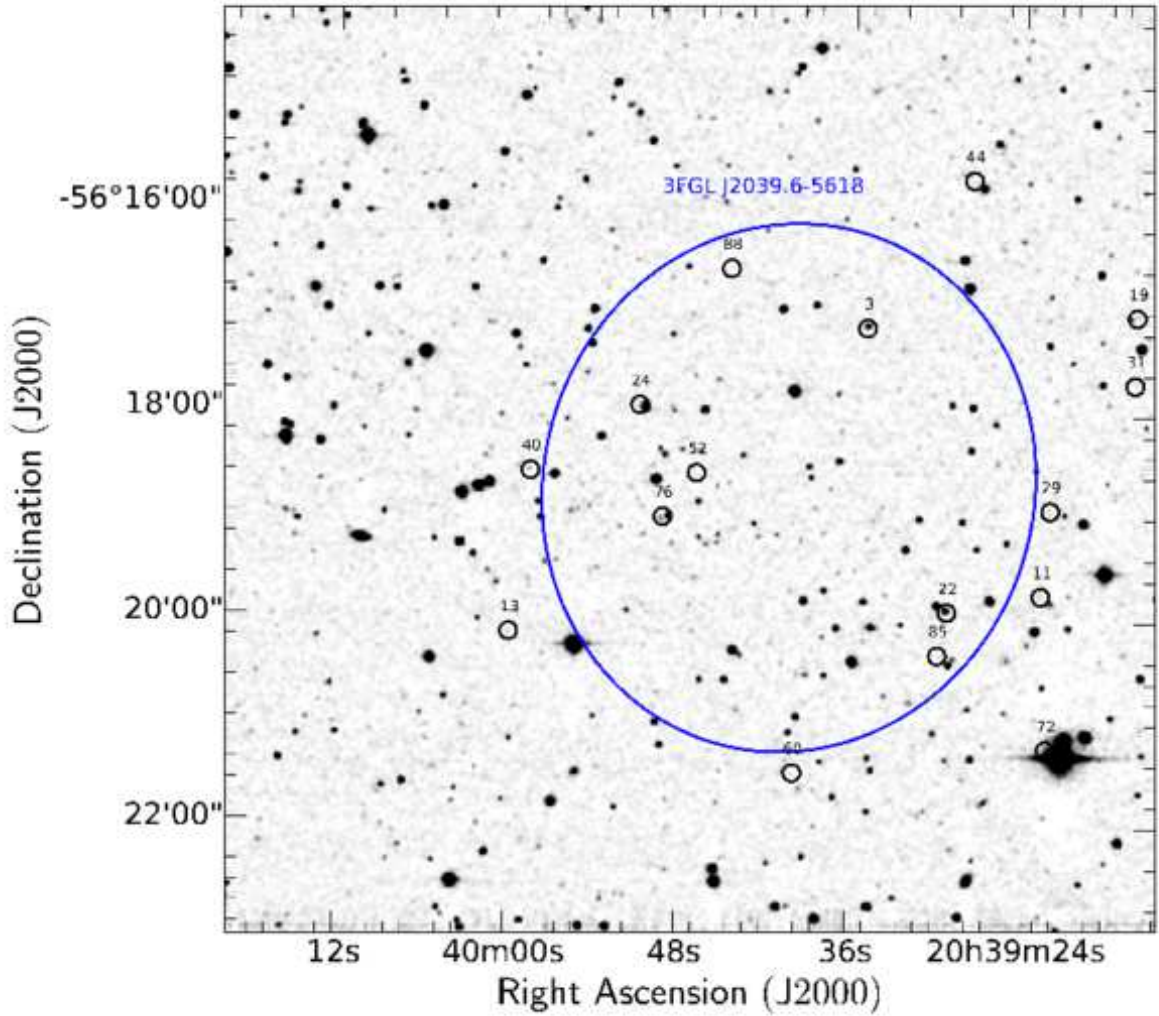}
\caption{GROND J-band image of the \msp\ field. The black circles indicate the positions of the \xmm\ sources detected within, or close to, the 3FGL error ellipse, here represented by the blue ellipse.  In all cases the circle radius has been arbitrarily set to 5\arcsec\ for a better visualisation. \label{grond-image}}
\end{figure}

We cross-matched the positions of all the 16 \xmm\ sources 
in Table 1 with the source catalogues obtained from the GROND observations.
We performed the source detection  on the single-band GROND exposures using the {\tt starfind} tool in {\sc IRAF}, matched the source catalogs over the different observations,  
	 and checked for variable sources against the median of all observations. Object photometry was computed using the task {\tt daophot} in {\sc IRAF} and the airmass correction was applied using the standard atmospheric extinction coefficients for the La Silla Observatory.
For the cross-match we used a radius obtained by combining the statistical $1 \sigma$ uncertainty on the X-ray source centroid plus  the 90\% confidence level systematic error associated with the absolute accuracy of the \xmm\ aspect solution, which is 1\farcs5 per coordinate\footnote{Calibration technical note XMM-SOC-CAL-TN-0018}. The uncertainty on the absolute astrometry of the GROND images (0\farcs3)
is much lower than the \xmm\ one,  and is accounted for by our choice of the matching radius.   
There are only six \xmm\ sources (Source 3, 13, 22, 24, 40, 76) with at least a candidate optical or near-IR counterpart in the GROND data (Figure \ref{grond-image}).

We also cross-correlated the \xmm\ source list  with the VISTA, OM, and UVOT source catalogues.   The source detection and photometry in the VISTA images 
were carried out as a part of the CASU pipeline (Sectn.~\ref{opt_uv}). For the OM images, the source detection 
and photometry were carried out 
with the SAS tasks {\tt omdetect} and
{\tt ommag}, respectively, 
and for the UVOT images
with the HEASOFT taks {\tt uvotdetect} and {\tt uvotsource}.
As before, the matching radius accounted for the uncertainty on the absolute astrometry of the used source catalogues, i.e.  $\la0\farcs3$  (VISTA; Emerson et al.\ 2004), 0\farcs5  (UVOT; Breeveld et al.\ 2010), and 0\farcs7  (OM; Page et al.\ 2012).
The cross-correlations with these catalogues provided additional near-IR, and near-UV magnitudes for some of the GROND sources and unveiled candidate counterparts for Source 11, 19, 29, 60, 88. We found no 
counterparts in the OM/UVM2 filter, whereas only Source 3 was detected in the UVOT/UVW2 images.
The optical, near-UV, and near-IR magnitudes of the candidate counterparts to the \xmm\ sources are summarised in Table 2.  
Only for the candidate counterparts to Source 3, 40, and 76 we have an adequate spectral coverage in at least the optical and near-IR.

\subsubsection{Variability analysis}\label{opt_var}

\begin{figure}
\includegraphics[width=14cm]{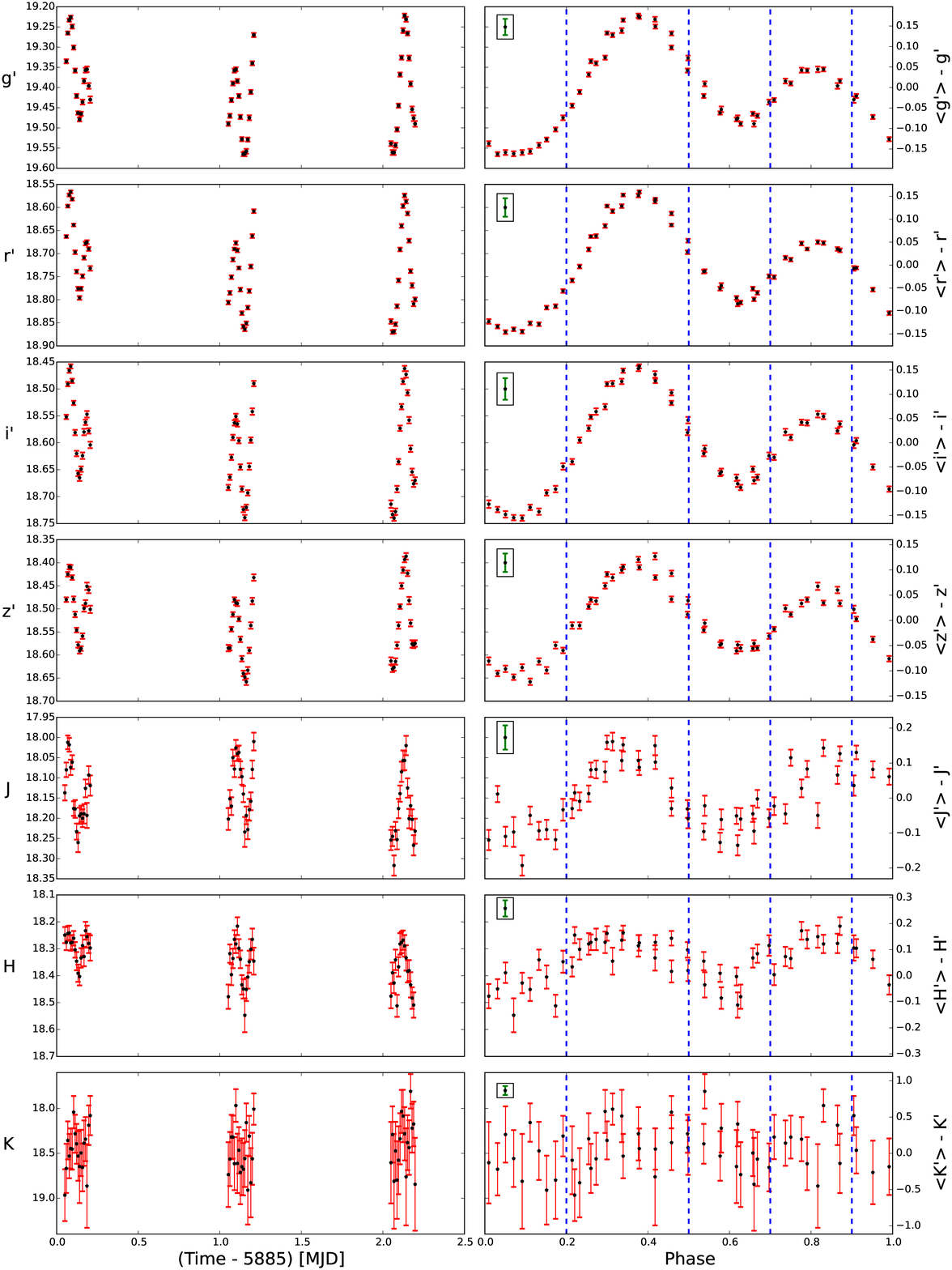}
\caption{{\em Left}:  Multi-band light curves of the optical counterpart to  \xmm\ Source 3.
{\em Right}: Light curves folded at the best-fitting period.  The axis on the right are magnitudes relative to the mean. The oscillations in magnitude in the same phase bins in the i' and z' bands are likely due to fringing.
Only statistical errors are plotted. The vertical ticks are the systematic errors associated with the accuracy of the photometric calibration (Sectn.~\ref{opt_uv}). The vertical dashed lines define the main peak ($\phi$=0.2--0.5), the secondary peak  ($\phi$=0.7--0.9), the ``bridge'' ($\phi$=0.5--0.7), and the ``off-peak'' ($\phi$=0.0--0.2 and $\phi$=0.9--1.0) regions. 
 \label{grond-lc}}
\end{figure}

Among the six \xmm\ sources with a possible GROND counterpart, only Source 3 is associated with a clearly variable object
($\chi^2=1232$, with 51 d.o.f.).
 Figure \ref{grond-lc} (left) shows the multi-band light curves of this object for the three nights spanned by the GROND observations. As seen, the light curves seem modulated, with an amplitude of $\sim$0.4 magnitudes in the g$'$ band. In particular, the modulation seems to be periodic and feature a double-peaked profile (night 1) with the two peaks only partially seen in night 2 and 3, likely owing to the different sampling of the light curve. This modulation is also seen in the r$'$, i$'$, and z$'$ light curves, with both shape and amplitudes similar to the g$'$ one,
  and is also recognised in the J and H-band light curves, and very marginally in the K-band one.
 This suggests that the observed modulation is real and not due, for instance, to possible problems with the photometry in a given filter.   As a check, for all filters we extracted the light curves of several stars of comparable brightness detected in the GROND FoV but found no evidence of such a modulation. This confirms that it is not due to random effects, such as variations in the sky conditions (transparency, background, moon illumination), or systematic effects, such as variations in the encircled flux due to the fixed size of the photometry aperture with respect to the seeing disk.  Furthermore, the fact that the modulation seems to be periodic argues against the possibility that it is produced by any of such effects and implies that 
  is associated with an intrinsic star variability. 

We computed the probability that the association between Source 3 and its candidate GROND counterpart is due to a chance coincidence. We computed the probability as $P=1-\exp(-\pi\rho r^2)$, where  $r$ is  the matching radius used for Source 3 (1\farcs8) and $\rho$ is  the  density of  stellar  objects  in the GROND field, regardless of their brightness, measured in the co-addition of all g'-band exposures. For a stellar density $\rho\sim0.0019$ square degree$^{-1}$ we estimated a chance coincidence probability $P\sim0.02$, which makes a chance coincidence unlikely.

To confirm the existence of a periodic modulation of the Source 3 candidate counterpart, we carried out a periodicity analysis
based on the Generalized Lombe-Scargle periodogram method (Lomb 1976; Scargle 1982; Zechmeister \& Kuerted 2009). For cross-checking purpose, we also used the phase dispersion minimisation technique (PDM; Stellingwerf 1978) using the {\tt pdm} code in {\sc IRAF} and the {\tt Period04} software package (Lenz \& Breger 2005), which is especially dedicated to the statistical analysis of large astronomical time series containing gaps
(see Figure \ref{grond-lc}). All methods indicate the presence of a periodicity with a period of $\sim$ 0.227 d. The analysis of the power spectrum of the optical time series shows a clear peak at the corresponding frequency, with a probability that is due to a chance noise fluctuation (false-alarm probability) of $\sim$ 1.8$\times$10$^{-15}$ ($\sim$ 8$\sigma$).  The best period was found at 0.22748$\pm$0.00043 d in the optical bands and at 0.22799$\pm$0.00062 d in the near-IR bands, where we estimated the period uncertainty following Gilliland et al.\ (1987). 
The optical and near-IR best-fit periods of the Source 3 candidate counterpart are consistent within the uncertainties,
which provides further evidence that the observed modulation is real.

The period of the optical/IR modulations seen in the GROND data for the Source 3 candidate counterpart 
is consistent with that seen in the \xmm\ data for Source 3 (0.2245$\pm$0.0081 d; Sect.~\ref{x_var}). Therefore, the detection of virtually the same periodicity clearly indicates that the two objects are associated.
Therefore, based upon the X-ray and optical variability at the same period and the relatively low chance coincidence probability, we regard the association between Source 3 and its GROND candidate counterpart as robust. Furthermore, since we based the selection of candidate X-ray counterparts to \msp\ on the search for variable sources, the optical/X-ray periodicity of Source 3 makes it a very promising candidate. Another possible candidate X-ray counterpart to \msp\  would be Source 11, the second brightest X-ray source detected in the 3FGL error circle, and the only other X-ray source with an obvious non-thermal spectrum (Table 1). However, Source 11 is not variable in X rays and is not associated with a periodic optical/near-IR GROND counterpart. Therefore, although it cannot be firmly ruled out as a candidate X-ray counterpart to \msp, as of now, it is a by far a less likely candidate than Source 3.  The same is true for all the other fainter X-ray sources in Table 1, whose poorly characterised X-ray spectra and sparse optical,  near-UV, and near-IR flux measurements hamper a detailed multi-wavelength analysis of their properties and a non-ambiguous classification.

\subsection{Characterization of the Source 3 counterpart}

\subsubsection{Folded light curves analysis}\label{opt_lc_src3}

\begin{figure}
\plotone{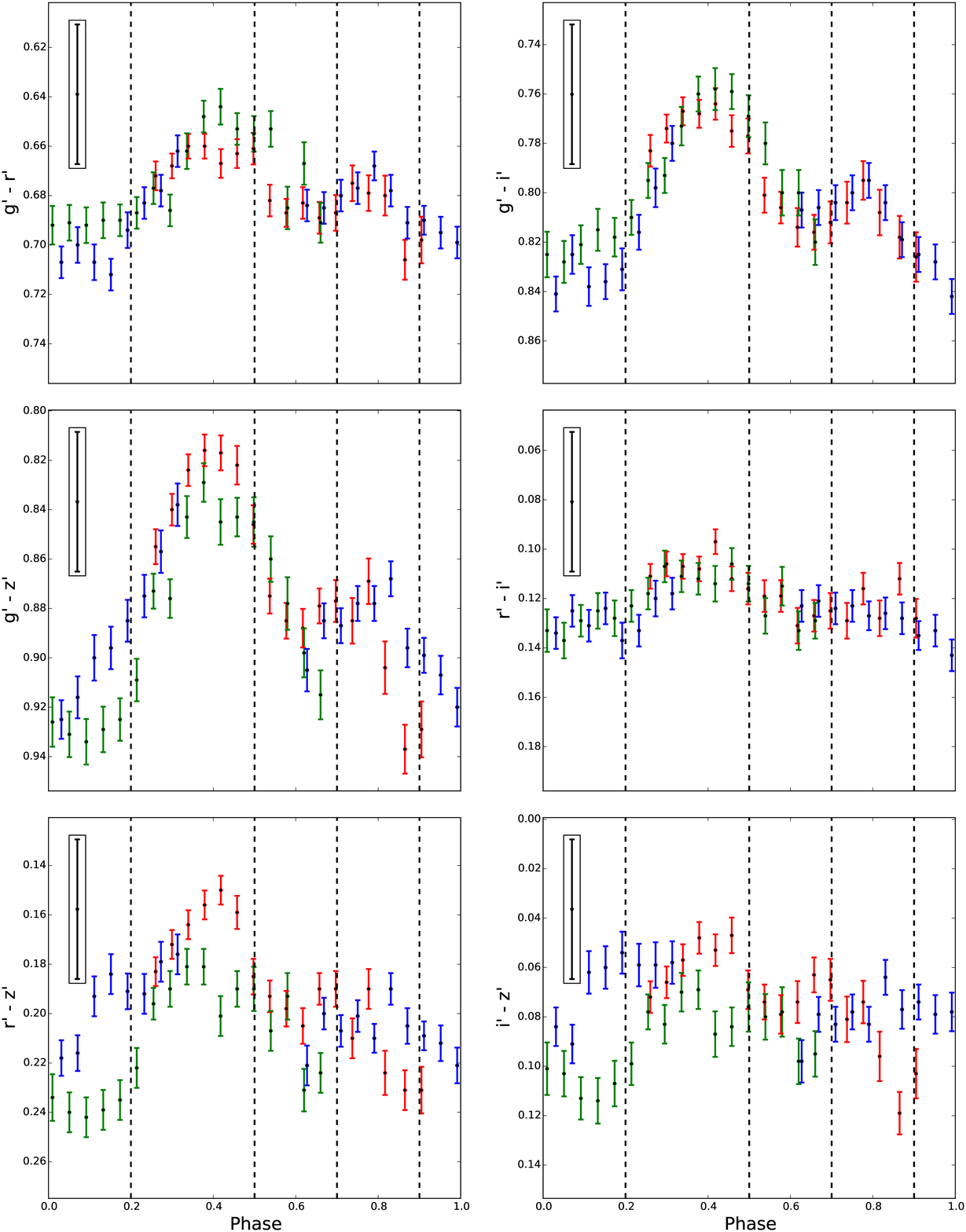}
\caption{
Colors of the Source 3 counterpart as a function of phase. Only statistical errors are plotted. The vertical ticks are the systematic errors in the color determination associated with the accuracy of the photometric calibration (Sectn.~\ref{opt_uv}). Different colors correspond to different nights, i.e. night 1 (red), night 2 (blue), night 3 (green).  The vertical dashed lines corresponds to the four regions defined in Figure \ref{grond-lc}. \label{grond-lc-diff}}
\end{figure}

The optical/near-IR  light curves of the Source 3 counterpart folded at the corresponding best-fit periods are shown in Figure \ref{grond-lc} (right).
A double-peak light curve is clearly visible, with a main peak and a secondary peak, separated in phase by $\sim 0.5$.
Per each band, 
we determined the phase of the two peaks by fitting a Gaussian function to their profiles 
to precisely compute their phase separations and errors. %
The peak phase separation remains constant in the optical ($0.436 \pm 0.003$; g'), while it seems to slightly increase in the near-IR ($0.517\pm 0.012$; J).  
To better recognise the light curve evolution, we defined four regions: the main peak ($\phi$=0.2--0.5), the secondary peak ($\phi$=0.7--0.9), the ``bridge'' ($\phi$=0.5--0.7), and the ``off-peak'' ($\phi$=0.0--0.2 and $\phi$=0.9--1.0).
The shape of the light curve is similar in all bands, 
but the profile of the modulation changes 
from the optical to the near-IR
(Figure \ref{grond-lc-diff}, left).
In particular, the primary peak  becomes broader 
and its amplitude
decreases, from $\sim$ 0.4 magnitudes in the g$'$ band 
to  $\sim$ 0.2 magnitudes in the H band, and the difference between the primary and the secondary peaks decrease from 0.133 magnitudes in the g' band to $\sim 0$ in the H band. Similarly, the amplitude of the "bridge" decreases, 
 becoming comparable to that of the "off-peak" region.
There is also a possible evidence of a colour variation
as a function of phase (Figure \ref{grond-lc-diff}, right), which might indicate that we are observing regions of the star surface at different temperatures. In particular, the colours seem to be bluer in coincidence with the two peaks, and bluer in coincidence with the primary peak than with secondary one. The colours also seem bluer
in the "bridge" than in the "off-peak" region.  The colour variation seems less marked at longer wavelengths 
 consistently with the fact that the light curve variations are smoother.
 Unfortunately, the  errors on the GROND photometry calibration (Sectn.~\ref{opt_uv}) make difficult to quantify the observed colour variations. 

We tried to align in phase the X-ray and optical light curves of Source 3 to check the relative alignments between  the  X-ray and optical peaks. However, the  difference between the epochs of the \xmm\ and GROND observations (MJD=56885--56887 and MJD=56575, respectively) corresponds to a maximum time difference of 312 d.  The accuracy on the best-fit optical period derived from the GROND observations is 0.00043 d (Sectn.~\ref{opt_var}), which corresponds to a phase uncertainty of $\sim 0.0019$. Thus, building a phase-coherent solution for the optical light curve backward to the epoch of the \xmm\ observation would bring an uncertainty of $\sim$ 0.59 on the absolute phase determination, which is larger than the phase separation between the two peaks ($\sim 0.3$ in the X rays and $\sim 0.45$ in the optical).

\subsubsection{Colour-magnitude analysis}\label{cm}

We computed the time-average multi-band photometry of the Source 3 counterpart from the GROND data
and obtained g$'$=19.40$\pm$0.02, r$'$=18.71$\pm$0.02, i$'$=18.59$\pm$0.02, z$'$=18.52$\pm$0.02,
J=18.13$\pm$0.03, H=18.33$\pm$0.03, K=18.35$\pm$0.06,
with all magnitudes in the AB system. The photometry errors account for both statistical errors and for the systematic uncertainty on the photometry calibration.  %
We also identified the Source 3 counterpart in the U and UVW1-filter exposures from the \xmm/OM 
with AB magnitudes  U=21.26$\pm$0.26 and m$_{UVW1}$=21.88$\pm$0.4 (Table 2), whereas it is not detected in the UVM2 filter down to a 3 $\sigma$ limiting magnitude of 21.99
and in the \sw/UVOT UVW2 image 
down to  
a limiting magnitude of 
23.34 (AB).
In the near-IR, we also identified the Source 3 counterpart in the VISTA images,
with AB magnitudes J=18.24$\pm$0.03, H=18.24$\pm$0.04, and K$_{\rm s}$=18.64$\pm$0.09,
converted from the native Vega survey system\footnote{http://casu.ast.cam.ac.uk/surveys-projects/vista/technical/filter-set}.
We found that the VISTA magnitudes (epoch 2010.6)  are all compatible with the GROND ones (epoch 2014.7),
after accounting for the difference between the K and K$_{\rm s}$ filters, which excludes long-term variability on year time scales from the Source 3 counterpart.

\begin{figure}
\includegraphics[width=9cm]{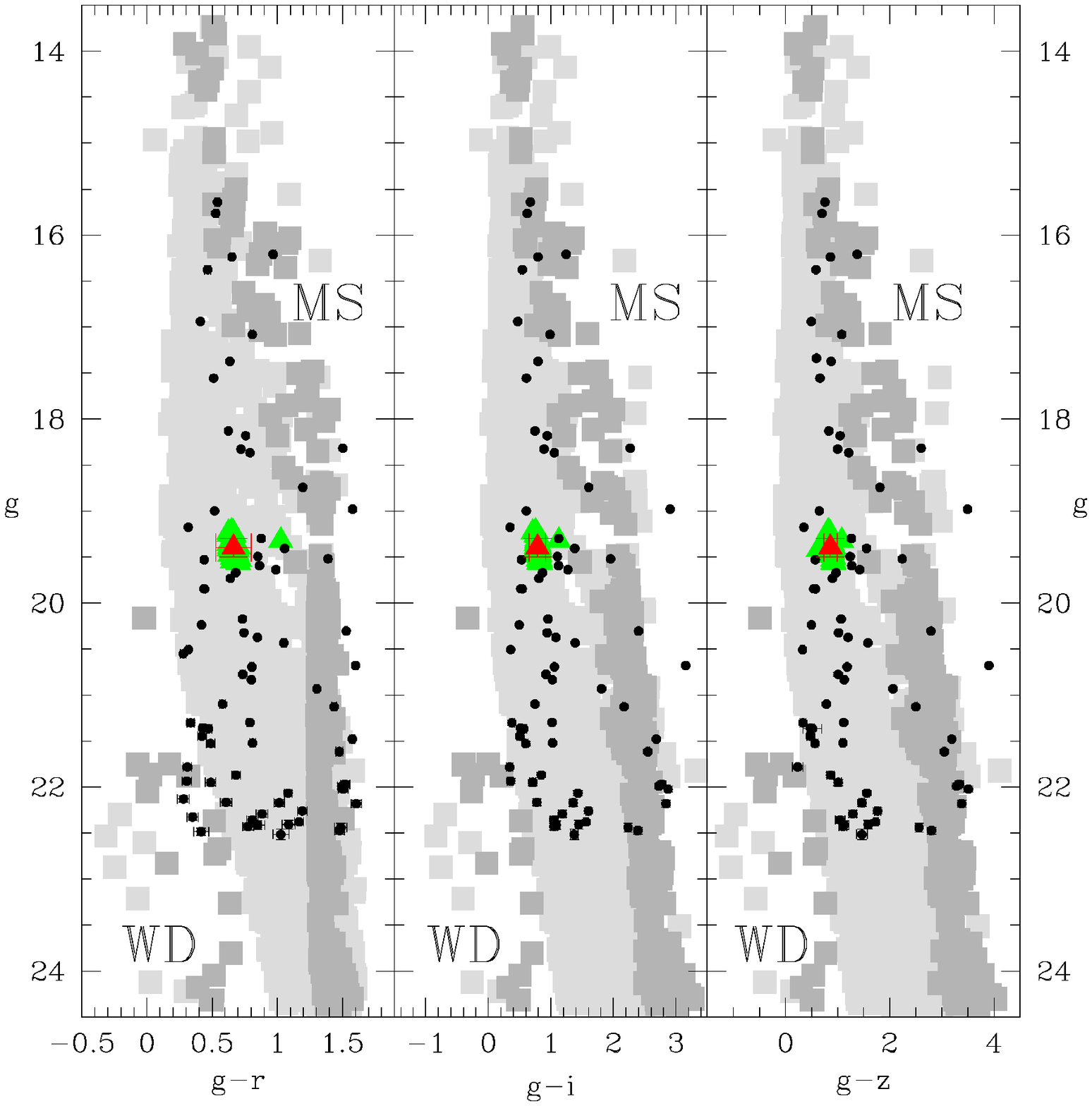}
\includegraphics[width=9cm]{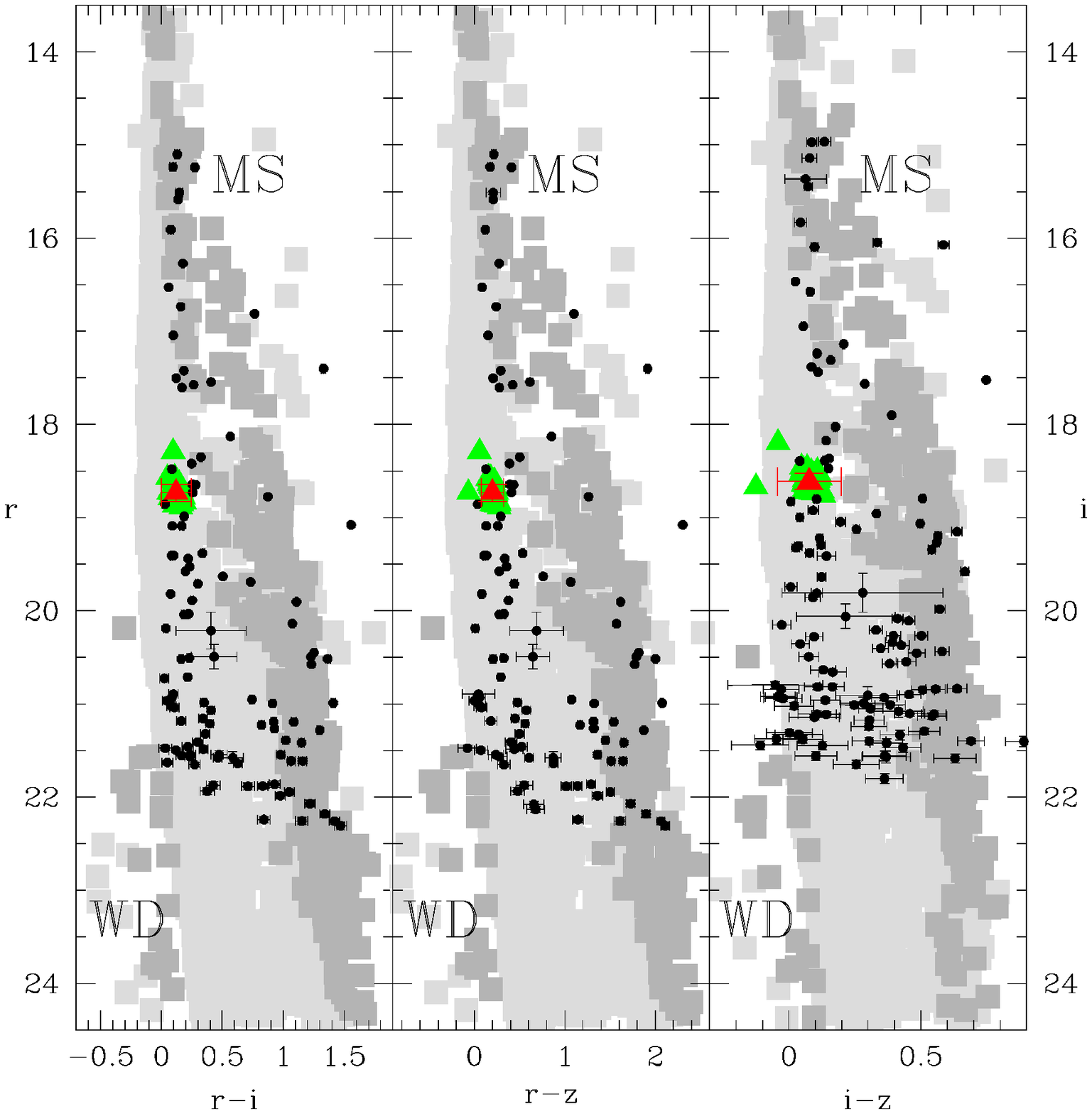}
\includegraphics[width=9cm]{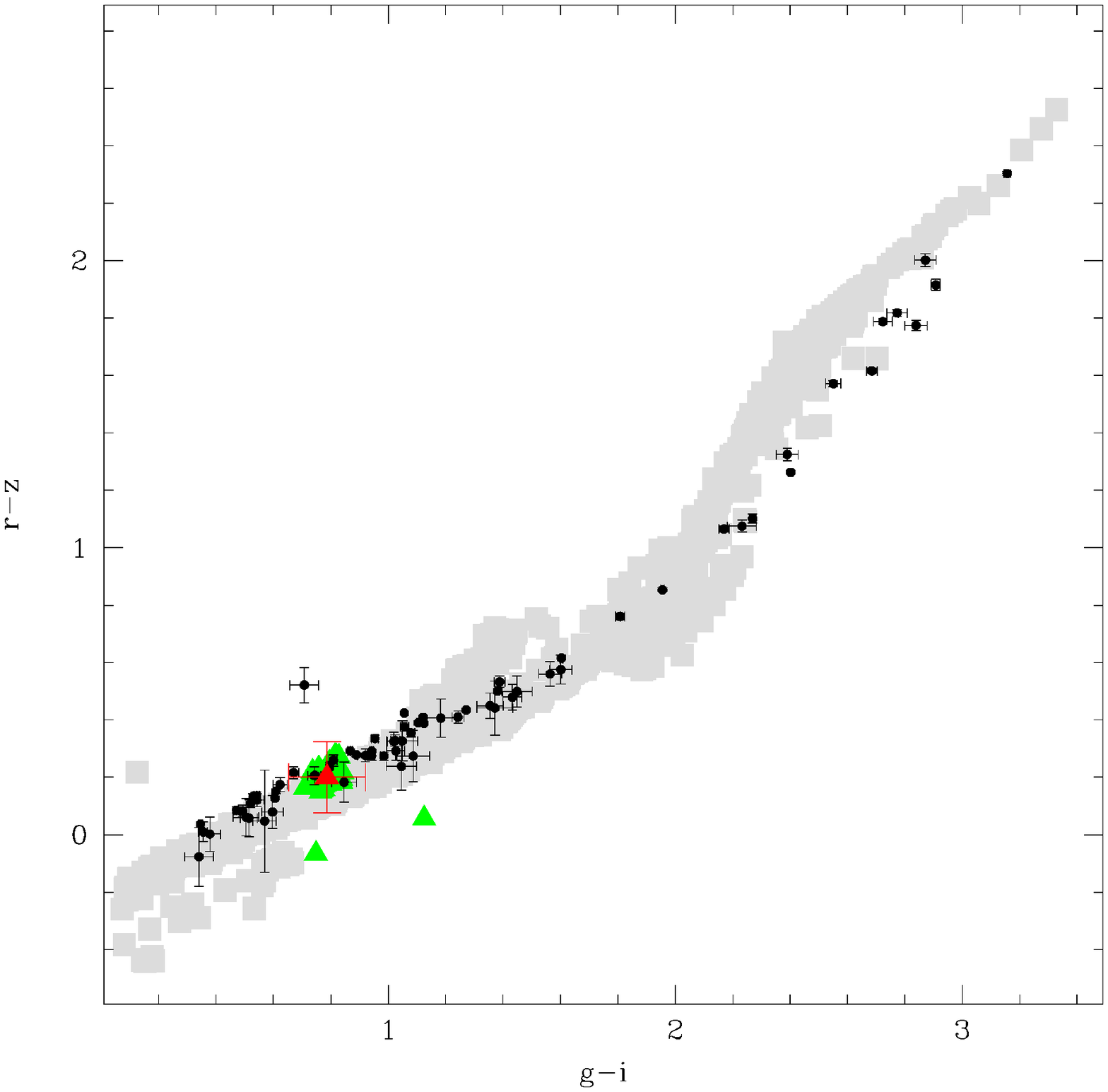}
\caption{{\em Top}: Observed CMDs for the \msp\ field obtained from the GROND time-averaged photometry.  {\em Bottom}:  Observed CC diagram. In all panels, the location of field stars is indicated by the black filled circles, whereas that of the optical counterpart of the \xmm\ Source 3 is indicated by the red filled triangle.   The filled green triangles indicates the counterpart location computed from the photometry computed on the single image.  Stellar sequences simulated from the Besan\c{c}on models for different values of distance are shown in light and dark grey. In the CM diagrams the dark grey regions correspond to distance values $200<d<900$ pc, whereas in the CC diagram they correspond to magnitudes within $\pm$ 0.05 the g'-band magnitude of the Source 3 counterpart.  The MS and WD branches are labelled. \label{grond-cmd}}
\end{figure}

We used the time-averaged g$'$, r$'$, i$'$, z$'$-band magnitudes of the Source 3 counterpart as a reference for its classification by analysing its location in the observed (i.e. not corrected for the reddening) colour-magnitude (CM) and colour-colour (CC) diagrams of the field. The diagrams are shown in Figure \ref{grond-cmd}, where the location of the field stars is shown by the black filled circles and that of the Source 3 counterpart as a red  filled triangle. Red and green triangles correspond to the location computed from the single-image photometry. 
In order to reject outliers and include only high-confidence measurements, we plotted only field stars for which at least 20 measurements per filter were available and with $\sigma < 0.08$.  We compared the observed CM and CC diagrams with simulated stellar sequences computed from the Besan\c{c}on models (Robin et al.\ 2004) for different stellar populations and distance values up  to 15 kpc. The simulated sequences are shown in Figure \ref{grond-cmd} as the grey scale map. The dark grey regions in the CM diagrams correspond to a likely distance range ($200 \la d \la 900$ pc) for Source 3, whereas in the CC diagram the dark grey region corresponds to simulated magnitudes that are within $\pm$ 0.05 the g$'$-band magnitude of the Source 3 counterpart.  

The distance to Source 3 is unknown a priori. The upper limit on the hydrogen column density derived from the fits to the X-ray spectrum of Source 3 ($N_{\rm H}<4 \times 10^{20}$ cm $^{-2}$; Table \ref{tab1}) indicates a distance lower than $\approx 0.9$ kpc, assuming the relation between distance and $N_{\rm H}$ of He, Ng \& Kaspi (2013). Without a parallax measurement, the distance to Source 3 cannot be precisely constrained from kinematic measurements of its optical counterpart.
The 
NOMAD  catalogue (Zacharias et al.\ 2005) gives a proper motion of $\mu_{\alpha} cos(\delta) =14\pm4$ mas yr$^{-1}$ and $\mu_{\delta}=-16\pm9$ mas yr$^{-1}$ in right ascension and declination, respectively.
This corresponds to a spatial velocity of $101^{+32}_{-30}$ $\times$ (d/1 kpc) km s$^{-1}$. If we equate it to the median of the transverse velocity distribution of MSPs ($\sim$108 km s$^{-1}$, with a standard deviation of $\sim$86 km s$^{-1}$)  computed from the Australia National Telescope Facility (ATNF) Pulsar Catalogue\footnote{{\tt http://www.atnf.csiro.au/people/pulsar/psrcat}}  (Manchester et al.\ 2005) we obtain a distance of $\sim$770--1500 pc, compatible with the estimate inferred from the $N_{\rm H}$.
Any determination of a lower limit on the distance is more uncertain. Again, if Source 3 were a binary MSP, 
the distance distribution of known binary MSPs
from the ATNF Pulsar Catalogue 
gives a probability of $\sim$ 0.006 to find one within a radius of 0.2 kpc. Thus, the assumed distance range  for Source 3 ($200 \la d \la 900$ pc) is reasonable. 

We  also used the upper limit on the $N_{\rm H}$ to infer an interstellar extinction along the line of sight   $E(B-V)<0.072$, after applying the relation of Predehl \& Schmitt (1995). Then, we computed the extinction in the different filters using the extinction coefficients of Fitzpatrick (1999).  The reddening vectors are shown in in Figure \ref{grond-cmd} for the limit case  $E(B-V)=0.072$.
	 Since the field stars are affected by an unknown interstellar extinction, and the simulations based on the Besan\c{c}on models simply compute a reddening scaled proportionally to the assumed distance in a given direction, introducing a reddening correction in our simulations might bias a direct comparison between the observed and the simulated stellar sequences.  Therefore, for simplicity, in all cases we simulated the stellar sequences assuming a null reddening.  Then, we used the reddening vectors as a reference to trace the extinction-corrected locations of the observed points for the Source 3 counterpart  (red points) along the simulated stellar sequences. 
As seen, the location of the Source 3 counterpart in the diagrams falls between the regions of the simulated MS and WD stellar sequences. Only for distances as low as $\sim$ 0.1 kpc the counterpart location in the diagrams could be compatible with a WD. However, if Source 3 is an MSP such possibility is unlikely (see above). Thus, we conclude that the star is, most likely, not a WD.

\subsubsection{Spectral analysis}\label{opt_spec}

We built the optical/near-UV/near-IR spectrum of the Source 3 counterpart using the available multi-band photometry (Sect.~\ref{cm}). In all cases, we used as a reference the measured AB magnitudes to compute the spectral fluxes at the filter peak wavelengths. As a reference for the interstellar extinction correction we used a maximum extinction value of $E(B-V)=0.072$, derived from the upper limit on the hydrogen column density $N_{\rm H}$ (Predehl \& Schmitt 1995) obtained with the fit with a PL spectral model (Sect.~\ref{spectra}). 
 
 We fitted the spectrum with both a single and a double BB spectral model. However, we found that the optical/near-UV/near-IR  data cannot be simultaneously fitted by a single BB and that a double BB is required to fit the entire spectrum ($\chi^2=20.6$, 6 d.o.f.). The inferred temperatures are $T_{\rm H} \sim 3700$ K for the hotter BB, which fits the optical/UV fluxes, and  $T_{\rm C} \sim 1600$ K, for the colder one, which fits the near-IR part of the spectrum. The two BB model is also consistent with the upper limits obtained in the OM/UVW2 and UVOT/UVM2 filters. The overall spectral shape and spectral parameters do not change significantly when adding the correction for the maximum estimated interstellar extinction ($T_{\rm H} \sim 3800$ K, $T_{\rm C} \sim 1600$ K). 
 While the temperature of the hot BB is compatible with the surface temperature of  a mid-MS companion star, the temperature of the colder one is too low to be entirely ascribed to the emission from the star surface. Although this can contribute to part of the observed near-IR emission, as indicated by the periodic modulation of the J, H and K-band light curves, an additional source external to the star is required to account for the low temperature of the BB that fits the near-IR part of the spectrum.  This source might be associated with emission from cold intrabinary gas or dust, maybe the residual of an accretion disk left over by a past phase of matter accretion on the neutron star. 
 
The analysis of the counterpart colours as a function of phase (Figure \ref{grond-lc-diff}, right) suggests that its spectrum slightly changes along the $\sim$ 0.2245 d period. To quantify this possible evolution we fitted the multi-band spectrum in the four phase intervals defined in Sect.~\ref{opt_lc_src3}.
Like in our phase-resolved colour analysis, we cannot use the single-epoch flux measurements in the U and UVW1 bands obtained with the  \xmm/OM. 
As done above, we fitted the four phase-resolved spectra using a two-BB model, considering both null and maximum interstellar extinction.
However, we did not find evidence of a significant spectrum evolution across the different phase intervals. This is partially ascribed to the fact the spectra are less constrained at shorter wavelengths without the flux measurements in the U and UVW1 bands.

\section{Discussion}

The optical and X-ray emission of Source $3$, modulated at a common periodicity of $\sim 0.2245\,{\rm d}$ is likely associated with the orbital motion of a tight binary
system. The IR/optical/UV spectrum points at a late spectral type (K or M) for at least one of the stars in the system (Secn.~\ref{opt_spec}). 
The other object may either be another non-degenerate star or a compact object, likely an MSP. In the first scenario, the X-ray activity, the orbital period and the shape of the optical light curves indicate the system could either be of the W UMa or $\beta$ Lyr type (Geske et al.\ 2006). However, the system would be extremely peculiar even for these classes of binaries. The orbital period would be one of the shortest ever observed; the spectral type one of the latest; the asymmetry and separation of the peaks in the optical light curves hard to explain.
 The second scenario, sounds more plausible.  Moreover, when compared to the $\gamma$-ray flux of \msp, $F_{\gamma} = (1.71 \pm 0.14) \times 10^{-11}$ erg cm$^{-2}$ s$^{-1}$, the 0.3--10 keV unabsorbed X-ray flux of Source 3 (F$_{\rm X} = 10.19^{+0.87}_{-0.82}\times10^{-14}$ erg cm$^{-2}$ s$^{-1}$) would give a $\gamma$--to--X-ray flux ratio $F_{\gamma}/F_X \approx 170$, which is consistent with that of MSPs  (Abdo et al.\ 2013; Marelli et al.\ 2015).  \msp\ is also  at a relatively high Galactic latitude, with $l=341.23\degr$ and
$b=-37.15\degr$, like most MSPs.  We note that the NOMAD proper motion of Source 3 in Galactic coordinates, $\mu_l =-17\pm4$ mas yr$^{-1}$ and $\mu_b=-12\pm9$ mas yr$^{-1}$, would suggest that it is moving towards the Galactic centre from its present location.
This, however, would not argue against an MSP identification since several MSPs in the ATNF catalogue have a negative proper motion in Galactic latitude.  Being older than a Gyr, MSPs are indeed expected to orbit in the Galactic potential and periodically move away and towards the plane. 
Interestingly, if ascribed to an orbital motion, the period of the observed optical flux modulation ($\sim$ 0.2245 d) would be comparable to
the orbital periods of BW and RB MSPs, which are usually less than one day (see, e.g. Roberts 2013).  
Since most binary MSPs detected as $\gamma$-ray pulsars by the {\em Fermi}-LAT are either BWs or RBs, the possible identification of Source 3 as a BW/RB system would, then, concur to make it the most likely X-ray counterpart to \msp.

The X-ray emission model for BW and RB is generally described by a the combination of a thermal component, which originates from the neutron star surface, and a non-thermal component, which primarily originates from an intrabinary shock and from the neutron star magnetosphere (Gentile et al.\ 2014).
The X-ray spectrum of Source 3 is predominantly non thermal, with a photon index $\Gamma_X=1.36$, and its X-ray luminosity in the 0.3--10 keV energy band is $L_X\sim10^{31}$~erg~s$^{-1}$~$d_{kpc}^2$, where $d_{kpc}$ is the distance in units of kpc. Both the X-ray luminosity and photon index of Source 3 are in general agreement with those of RB/BW MSPs (Gentile et al.\ 2014; Roberts 2014), although its relatively hard X-ray spectrum would point more at a RB than a BW. 
 As we noted in Sectn.\ref{spectra},  a thermal component might be present in the X-ray spectrum of Source 3. However, 
further observations are necessary to clearly discriminate between a purely non-thermal and a composite spectral model.

The X-ray light curve of Source 3 can be explained assuming a binary MSP scenario,
the emission from the intrabinary shock is expected to be modulated at the orbital period. Recent studies suggest that the X-ray modulation may be due to synchrotron beaming, Doppler boosting of the flow within the shock, or obscuration by the companion (Bogdanov et al.\ 2011; Gentile et al.\ 2014; Roberts 2014). The shape of the X-ray light curve strongly  depends on the geometrical and physical parameters of the binary system but, on average, it is characterized  by 
an overall increase of a factor of $\sim2$--3 around inferior conjunction. The light curves of many BWs and RBs show a double-peaked structure due to the obscuration of part of the shock by the companion around the superior conjunction. The inclination angle of the system, as well as the ratio of the companion radius to the intrabinary separation, characterize the phase separation and levels of the two peaks. The 
X-ray light curve of Source 3 clearly points at a BW/RB scenario. Its amplitude changes by a factor of $\sim3$ during the orbit, the minima 
occur at orbital phases 
$0.0$--0.1 and $0.9$--0.0, which corresponds to the superior conjunction, whereas the maxima occur at phase $0.4$--0.7, which corresponds to the inferior conjunction. 
Unfortunately, we cannot align in phase X-ray and optical light curves (Sectn~\ref{opt_lc_src3}) to confirm this scenario.

The optical emission in BWs and RBs is dominated by the non-degenerate companion and is characterized by significant luminosity and colour variations. The optical light curve profile 
can bring the signature of two different effects: the tidal distortion of the companion star surface,  due to the gravitational pull of the MSP, and the heating of the companion star surface, due to irradiation of relativistic photons and/or high energy particles from the MSP (Breton et al.\ 2013). 
Most BWs exhibit a single-peak optical light curve (e.g., Stappers  et al.\ 2001), due to the irradiation of the nearly fully peeled 
companion star from the MSP.
Half of the RBs also feature a single peak,  whereas the rest
feature a two-peaked light curve, with the peaks occurring at specific orbital phases due to the viewing geometry of the tidally-distorted, non-degenerate, companion star. 
In the latter case, the companion star nearly fills its Roche lobes and 
tidal distortion effects dominate over the heating effects.
Examples of RBs with single-peak light curves are: PSR\, J1023+0038 (Thorstensen \& Armstrong 2005), J2215+5135 (Schroeder \& Halpern 2014), J2339$-$0533 (Romani \& Shaw 2011), and J1227$-$4853 (Bassa et al.\ 2014), while examples of RBs with double-peak light curves are: PSR\, J1628$-$3205  (Li et al.\ 2014), J1723$-$2837 (Van Staden et al.\ 2013), J1816+4510 (Kaplan et al.\ 2012), and J2129$-$0428 (Hui et al.\ 2015). The double-peak light curves of the RBs look very similar to that of Source 3, displaying asymmetries in the relative phase of the 
two peaks. In particular, the optical light curve of PSR\,  J1628$-$3205  (Li et al.\ 2014) shows the same remarkable asymmetry between the two peaks, with a main and a secondary peak, as observed in the Source 3 counterpart.
This similarity would, then, support the hypothesis  that Source 3 is a RB. 

The $\gamma$-ray error ellipse of \msp\  has been observed four times at the CSIRO Parkes telescope searching for pulsations up to a $DM=200$ pc cm$^ {-3}$
(Camilo et al.\ 2015). No pulsations were found, but this is not unexpected, because RBs are very elusive targets in radio. In fact, the intrabinary material ablated from the star causes strong and variable scattering and absorption of radio waves. The radio detection of the pulsations often requires several dedicated long observations (Ray et al.\ 2013; Ray et al., in preparation).

\subsection{Modelling of the optical light curve}\label{opt_model}

Standard models for the RB optical light curves based on tidal distortion and pulsar irradiation (e.g., Thorstensen \& Armstrong 2005) cannot fit well neither the asymmetric peaks nor the peak separation. Therefore, we built a simple three-dimensional model including an additional component related to the asymmetric irradiation of the companion star to fit the light curve of the Source 3 counterpart.
In this process, we considered only the optical light curves because are those with the highest signal--to--noise. 

In order to probe the RB scenario and estimate its physical parameters we built a simple three-dimensional model of a RB binary system with very few free parameters  and fit it to the observed optical light curves of Source 3.  In this model,  the shape of the companion star is approximated by a sphere and a tangent cone pointing to the neutron star. The cone is meant to account for the tidal deformation of the star as it approaches filling its Roche lobe. By locking the star rotation to the orbital motion we assumed two different brightnesses for night and day, a characteristic commonly found in  RB systems.
The asymmetry in the two peaks of the optical light curve implies some asymmetry in the physical system that produces it. Therefore,  we allowed a tilt angle between the cone axis and the day/night separator line, as measured on the orbital plane. As for other RBs, we assumed a perfectly circular orbit, reducing the number of free parameters in the Kepler's orbital parameter space. To summarize, our model accounts for four geometrical parameters:  the star deformation (distance of the cone tip from the star center in star radius units); the night/day asymmetry (angle between the cone axis and the line of sight to the star);  the orbital inclination (angle between the line of sight and the orbital plane);  the epoch of quadrature (when the projected axis of companion star orbit lies perpendicular to the line of sight).
Furthermore, for each band we have two free parameters:  the light curve normalization and the brightness ratio between night and day.

We fitted our model to the data, including both statiscal and systematic errors, and found  good qualitative agreement for a narrow range of parameters. The overall goodness of the fit  turns out to be 197.3 (196 d.o.f.). The configuration obtained from the best fit to the model implies a deformation of 1.519$\pm$0.009 for the companion star, indicating that it is subject to strong tidal effects, and a large value of the asymmetry, $52.1^{\circ} \pm 1.2^{\circ}$.  The orbital inclination of the binary system obtained from the best fit is $48.9^{\circ} \pm 0.6^{\circ}$, whereas the epoch of quadrature is at MJD 56884.9667$\pm$0.0003. 
Finally, the best-fit brightness ratios between the night and day sides in the four bands are: 0.789$\pm$0.007 [g']; 0.828$\pm$0.007 [r']; 0.851$\pm$0.007 [i']; 0.878$\pm$0.008 [z'].

The model assumes, as a first approximation, that the night and day sides of the star are in local thermodynamical equilibrium at different temperatures T$_{\rm night}$ and T$_{\rm day}$, respectively.  A brightness ratio implies a relation between these two temperatures, as a function of the temperature itself. This relation weakly depends on the interstellar extinction, which can only be constrained by our upper limits on the $N_{\rm H}$. We computed these relations for each band, adding in quadrature the uncertainty associated with the interstellar extinction correction. The values of temperature ratios are compatible in the four bands, at 1$\sigma$, only for a value of the day-side temperature T$_{\rm day}<9900$ K. This constraint does not rely on the normalizations, hence on the accuracy of the photometric calibration.
It is a side product of our model, consistent with the spectral results obtained in Section 3.3.3.
On the other hand, the model implies a simple relation between the temperature ratio and their absolute value:

\begin{equation}
\frac{\textrm{T}_{\rm day} - \textrm{T}_{\rm night}}{\textrm{T}_{\rm day}} = 8.666\times10^{-3} \left(\frac{\textrm{T}_{\rm day}}{1000\ \textrm{K}}\right) - 1.890\times10^{-4} \left(\frac{\textrm{T}_{\rm day}}{1000\ \textrm{K}}\right)^2
\end{equation}

The large value of the asymmetry, $\sim 50^{\circ}$, is implied by the different levels of minima and maxima in the light curves. Similar features are observed in other RBs, like PSR\,  J1628$-$3205  (Li et al.\ 2014), where asymmetry was also proposed as an explanation. In those cases, a measure of this effect was not possible, using the standard modelling tools, which motivated our alternative modelling. In our model, an asymmetry causes a phase shift between the sinusoid that represents the day/night variation and the peaks due to the bump. If the brighter temperature is induced by irradiation from the neutron star, this must be channeled through wind rather than photons. In fact, high-energy radiation travels straight lines, implying a symmetric heating, while particles may follow other trajectories.
Alternatively, the apparent difference in temperature could be due to very large spots on the surface of the low-mass star. This would not be totally unexpected given the purely convective nature of this kind of stars, and the perturbation induced by the pulsar. If this were the case, a future observation may reveal a different asymmetry, indicating migration of the spots on the surface.

\section{Conclusions}

We carried out multi-wavelength observations of the unidentified {\em Fermi}-LAT source \msp\ with \xmm\ and GROND. We detected a likely X-ray counterpart (Source 3) within the $\gamma$-ray error box of \msp, which is characterised by a PL X-ray spectrum ($\Gamma_X =1.36\pm0.09$) that is indicative of strong magnetospheric emission. The upper limit on the hydrogen column density inferred from the X-ray spectral fit, $N_{\rm H} < 4 \times 10^{20}$ cm$^{-2}$, imply a distance probably lower than 1 kpc. The X-ray light curve of Source 3 features a modulation with a period of 0.2245$\pm$0.0081 d and two peaks, separated in phase by $\sim 0.3$.  Using the GROND data, we found an optical counterpart to Source 3 that features an asymmetric, double-peaked, light curve and a flux modulation with a period of 0.22748$\pm$0.00043 d, coincident with that measured in the X rays. If we interpret this periodicity as the orbital period of a compact binary system, Source 3 would probably be a binary MSP and, as such, a very likely counterpart to \msp. In particular, both the optical colors of its putative companion star and the optical light curve profile, with two asymmetric peaks separated in phase by $\sim 0.5$,  suggest that Source 3/\msp\ is a RB.  This hypothesis is supported by  the lack of  apparent radio emission, which would be explained by eclipse of the radio signal as it propagates through the evaporated atmosphere of the irradiated MSP companion, as observed in many RBs.

Optical spectroscopy observations of Source 3 will be essential to obtain a more accurate classification of its companion star  and measure its radial velocity curve, crucial to confirm the binary MSP scenario and determine the orbital  parameters of the binary system. A precise determination of the orbital parameters is necessary to fold $\gamma$-ray
photons and search for the MSP pulsations. A timing solution, that can be extended
back in time to the launch of {\em Fermi} in 2008, will provide even tighter constraints
on the orbital parameters and their evolution. This has a potentiially huge
scientific payoff in terms of fundamental physics (Romani et al., 2012; Pletsch \& Clark, 2015)

\acknowledgments
The research leading to these results has received funding from the European Commission Seventh Framework Programme (FP7/2007-2013) under grant agreement n. 267251. This work was supported by the ASI-INAF contract I/004/11/0,
art.22 L.240/2010 for the project ``Studio di sorgenti di alta energia con Swift''. 
CD acknowledges support through EXTraS, funded from the European
Union's Seventh Framework Programme for research, technological
development and demonstration under grant agreement no 607452.
Part of the funding for GROND (both hardware as well as personnel)
was generously granted from the Leibniz-Prize to Prof. G. Hasinger
(DFG grant HA 1850/28-1).

{\it Facilities:} \facility{XMM-Newton},  \facility{GROND}, \facility{Swift}, \facility{Suzaku}, \facility{VISTA}

\clearpage

\begin{landscape}
\begin{table}
\begin{center}
\caption{Spectral properties of the \xmm\ sources detected within, or close to, the error ellipse of  \msp. Spectral models are:  {\em power-law} (PL), {\em apec} (AP), and {\em black-body} (BB).  \label{tab1}}
\begin{tabular}{cccccccc}
\tableline\tableline
\multirow{2}{*}{Source ID} & J2000 coord. & Counts rate& Spectral & N$_\textrm{H}$ & $\Gamma_X$ & Flux$_{[0.3-10 \textrm{ keV}]}$ & 
Variability\\
  & RA Dec [$^{\circ}$] (stat. err.\tablenotemark{a})  & $10^{-3}$ cts/s & model & $10^{22}$ cm$^{-2}$ &  & $10^{-14}$\,erg cm$^{-2}$ s$^{-1}$  & $\chi^2$ (d.o.f.)\\
\tableline
3 & 309.8956 -56.2861 (0.3\arcsec) & 23.78$\pm$0.99 & PL & $<0.04$ & 1.36$\pm$0.09 & 10.19$^{+0.87}_{-0.82}$ & 66.18 (16)\\
11 & 309.8439 -56.3292 (0.6\arcsec) & 10.76$\pm$0.79 & PL & $<0.16$ & 1.33$^{+0.24}_{-0.17}$ & 5.73$^{+0.66}_{-0.63}$ & 3.92 (8)\\
13 & 309.9995 -56.3359 (0.5\arcsec) & 4.90$\pm$0.57 & PL/AP & $<0.26$ & 2.07$^{+0.55}_{-0.37}$ & 2.66$^{+0.87}_{-0.45}$  & 4.0 (8)\\
19 & 309.8170 -56.2838 (0.9\arcsec) & 6.51$\pm$0.72 & PL/AP & $<0.24$ & 1.54$^{+0.52}_{-0.22}$ & 2.98$^{+0.67}_{-0.60}$ & 3.8 (8)\\
22 & 309.8713 -56.3320 (0.7\arcsec) & 7.84$\pm$1.36 & PL/AP & $<0.77$ & 1.80$^{+0.74}_{-0.38}$ & 3.37$^{+1.84}_{-0.59}$ & 11.9 (8)\\
24 & 309.9619 -56.2989 (0.8\arcsec) & 4.44$\pm$0.46 & PL/AP & $<0.46$ & 1.86$^{+1.08}_{-0.37}$ & 1.24$^{+0.41}_{-0.31}$ & 4.1 (8)\\
29 & 309.8415 -56.3154 (1.0\arcsec) & 5.19$\pm$0.71 & PL/AP/BB & $<1.05$ & 2.37$^{+2.24}_{-0.70}$ & 1.36$^{+1.15}_{-0.44}$ & 6.2 (8)\\
31 & 309.8172 -56.2948 (0.8\arcsec) & 4.21$\pm$0.54 & PL/AP/BB & $<0.91$ & 1.80$^{+0.72}_{-0.58}$ & 3.31$^{+1.60}_{-0.68}$ & 7.4 (8)\\
40 & 309.9936 -56.3097 (1.1\arcsec) & 3.03$\pm$0.50 & PL/AP\tablenotemark{b} & 0.05 & 2 & 0.63$^{+0.23}_{-0.22}$ & 6.25 (6)\\
44 & 309.8651 -56.2619 (1.5\arcsec) & 2.46$\pm$0.44 & PL/AP/BB\tablenotemark{b} & 0.05 & 2 & 0.11$^{+0.18}_{-0.04}$  & 5.5 (5)\\
52 & 309.9451 -56.3099 (1.3\arcsec) & 2.00$\pm$0.35 & PL/BB & $<5.21$ & 1.11$^{+3.00}_{-0.77}$ & 1.67$^{+1.40}_{-0.70}$  & 2.6 (5)\\
60 & 309.9159 -56.3583 (1.4\arcsec) & 1.78$\pm$0.32 & PL/AP/BB\tablenotemark{b} & $<1.42$ & 2.33$^{+1.01}_{-0.83}$ & 0.49$^{+0.23}_{-0.25}$  & 4.4 (4)\\
72 & 309.8419 -56.3541 (1.6\arcsec) & 1.58$\pm$0.41 & PL/AP/BB\tablenotemark{b} & 0.05 & 2 & 0.74$^{+0.24}_{-0.23}$ & 0.7 (3)\\
76 & 309.9548 -56.3170 (2.0\arcsec) & 1.10$\pm$0.29 & PL/AP/BB\tablenotemark{b} & 0.05 & 2 & 0.68$^{+0.15}_{-0.21}$ & 1.0 (3)\\
85 & 309.8740 -56.3390 (1.6\arcsec) & 1.24$\pm$0.43 & PL/AP\tablenotemark{b} & 0.05 & 2 & 0.74$^{+0.26}_{-0.27}$ & 0.64 (3)\\
88 & 309.9357 -56.2766 (1.7\arcsec) & 1.62$\pm$0.43 & \nodata & \nodata & \nodata & \nodata  & 8.89 (3)\\
\tableline
\end{tabular}
\tablenotetext{a}{Here we report only the $1\sigma$ statistical error, the $1\sigma$ systematic error is about 1.5\arcsec\ for each X-ray source.}
\tablenotetext{b}{Owing to  the low number of counts, we fixed the photon index ($\Gamma_{\rm X}$) to 2 for a PL model and the temperature ($kT$) to 3.5 and 0.2 for an AP and a BB model respectively.}
\tablecomments{Results of the spectral analysis of the \xmm\ sources detected within an error circle of 1.5 times the 95\% confidence error ellipse of \msp. If the spectrum is well fitted by more than one model, we report only the PL parameters. Here, we report the best-fit X-ray position, the count rate
the best-fit spectral model, the best-fit column density, the best-fit photon index, 
the unabsorbed X-ray flux in the 0.3--10 keV energy band, and the variability test described in the text. The errors are at a 90\% confidence and the upper limits at $3\sigma$.}
\end{center}
\end{table}
\end{landscape}

\begin{landscape}
\begin{table}
\begin{center}
\caption{Magnitudes of the optical, near-IR, and near-UV counterparts to the \xmm\ sources detected within, close to, the error ellipse of  \msp\ (Table 1), derived from the GROND, VISTA, OM, and UVOT observations. For GROND we report the average magnitudes only. The matches have been computed using a radius of 3\arcsec\ which accounts for both the statistical and systematic errors on the \xmm\ source coordinates (see Table 1) and the accuracy of the absolute astrometry of the optical, near-IR, and near-UV images. All magnitudes are in the AB system. \label{tab2}}
\begin{tabular}{c|ccccccc|ccc|ccc|cccccc}
\tableline\tableline
ID & \multicolumn{7}{c}{GROND} & \multicolumn{3}{c}{VISTA} & \multicolumn{3}{c}{OM} & \multicolumn{1}{c}{UVOT} \\ \hline
      & g'         & r'            & i'         & z'          & J          & H         & K                  &  J      & H       & K$_{\rm s}$        & U & UVW1 & UVM2 &  UVW2  \\ \hline
 3   & 19.40 & 18.73 & 18.72 & 18.61 & 18.53 & 18.15 & 18.34 &18.24 & 18.24 & 18.64 &     & 21.26 &              & 21.88    \\
 11  &             &            &            &             &              &             &                        &20.33& 20.02 & 20.73 &      &          &              &             \\
 13  &             &             & 22.61&            &              &             &                      &          &           &     20.70    &         &   21.27 &           &                            \\
19 &             &             &             &            &              &             &                      &          &           &     20.64    &        &    &           &                            \\
22 &              &             &            &             & 19.37 &            &  17.88             &          &           &           &             &    &        &               \\
     &              &             &            &             & 18.02 & 17.96 &                         &17.84 & 17.93  &17.97           &     &           &             &               \\
24     & 24.21 &               &            &           &              &            &                      &          &           &         &         &    &           &                            \\
29     &             &               &            &           &               &           &                                &          &            &        & 20.83 & 20.39 & & \\
40  & 21.95&21.22    & 21.45 & 22.92&             &            &             &           &   19.82       & 19.85          & 20.10          & 22.08     &           &             &               \\
72   &            &             &             &            &              &             &                      &          &     19.66      &         &         &    &           &                            \\
60  &  &               &            &           &               &           &              &                             &            &        &  & 19.33 & & &  \\
 76 &19.64 & 18.36   &  18.63& 18.20& 17.65&17.66&17.77          &      17.63    &       17.58    &   17.90       &     &           &             &               \\
      &            &               &             &21.52&           &            &                     &  19.36        &18.84           &  18.64         &     &           &             &               \\
      &            &               &             & 23.97 &             &            &                        &          &           &           &     &           &             &               \\  
88  &             &              &              &          &             &            &                        &  20.93&  21.31  &           &     &           &             &               \\ 
\tableline
\end{tabular}
\end{center}
\end{table}
\end{landscape}

\begin{figure}
\includegraphics[width=12cm]{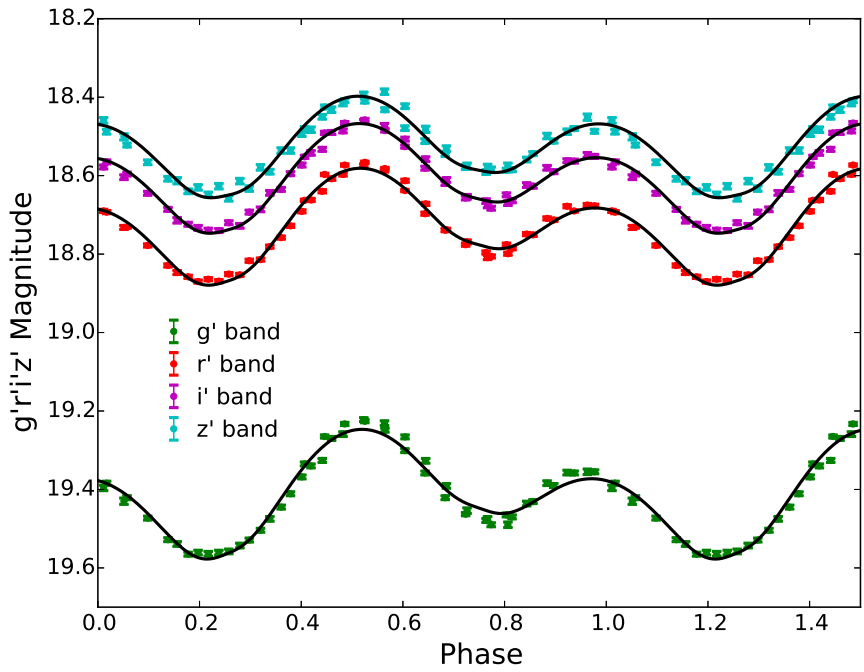}
\caption{Multi-band light curve of the optical counterpart of the \xmm\ Source 3. g$'$, r$'$, i$'$ and z$'$ bands are marked by green, red, magenta and cyan circles. Only statistical errors are plotted. The black lines display the best-fit light curve calculated using the model described in Sec.~\ref{opt_model}. \label{grond-lc-model}}
\end{figure}

\end{document}